\documentclass[10pt]{article}%
\usepackage[utf8]{inputenc}%
\usepackage[english]{babel}%
\usepackage[dvips]{graphicx}%
\usepackage{hhline}%
\usepackage{longtable}%
\usepackage{amsmath}%
\usepackage{amssymb}%
\usepackage{caption2}%
\usepackage{multicol}%
\renewcommand{\baselinestretch}{1.1}
\renewcommand{\topfraction}{1}
\renewcommand{\bottomfraction}{1}
\renewcommand{\thefootnote}{\ifcase\value{footnote}\or(*)\or
(**)\or(***)\or(****)\fi}
\begin{document}

\twocolumn[\protect{%
\begin{center}
{\bf \large LARGE-SCALE STRUCTURE AND GALAXY MOTIONS IN~THE~LEO/CANCER~CONSTELLATIONS}\\
\bigskip
{\large{}I.\,D.\,Karachentsev$^{1,\,}$\footnotemark,
O.\,G.\,Nasonova$^{1}$,
V.\,E.\,Karachentseva$^{2}$\\}
\bigskip
{\footnotesize\itshape$^{1}$Special Astrophysical Observatory of Russian AS, Nizhnij Arkhyz 369167, Russia\\
$^{2}$Main Astronomical Observatory, National Academy of Sciences of Ukraine, Kiev, 03680 Ukraine\\}
\bigskip
(Received November 14, 2014; Revised December 4, 2014)
\bigskip
\end{center}

\begin{quote}
In the region of the sky limited by the coordinates ${\rm
RA}=7^h\hspace{-0.4em}.\,0$--$12^h\hspace{-0.4em}.\,0$, ${\rm
Dec}=0^{\circ}$...$+20^{\circ}$ and extending from the Virgo Cluster to the
South Pole of the Local Supercluster, we consider the data on the galaxies with
radial velocities $V_{\rm LG}\lesssim 2000$~km/s. For 290 among them, we
determine individual distances and peculiar velocities. In this region, known as
the local velocity anomaly zone, there are 23 groups and 20 pairs of galaxies
for which the estimates of virial/orbital masses are obtained. A nearby group
around NGC\,3379~$=$ Leo\,I and NGC\,3627 as well as the Local Group show the
motion from the Local Void in the direction of Leo cloud with a characteristic
velocity of about $400$~km/s. Another rich group of galaxies around NGC\,3607
reveals peculiar velocity of about $-420$~km/s in the frame of reference related
with the cosmic background radiation. A peculiar scattered association of dwarf
galaxies Gemini Flock at a distance of $8$~Mpc has the radial velocity
dispersion of only $ 20$~km/s and the size of approximately $0.7$~Mpc. The
virial mass estimate for it is 300 times greater than the total stellar mass.
The ratio of the sum of virial masses of groups and pairs in the Leo/Can region
to the sum of stellar masses of the galaxies contained in them equals 26, which
is equivalent to the local average density $\Omega_m(\mbox{local}) = 0.074$,
which is 3--4 times smaller than the global average density of matter.
\end{quote}

{\bf{}Keywords}: galaxies: kinematics and dynamics---galaxies: distances and
redshifts---galaxies: groups}

\vspace{1cm}]

\footnotetext[1]{Electronic address: ikar@sao.ru}

\section{Introduction}

The modern cosmological paradigm assumes that the galaxy formation occurs in the
areas of concentration of dark matter, to where the baryonic matter is inflowing
and triggering the star formation processes. Within this concept, the apparent
distribution of galaxies follows the distribution of dark matter, but with a
somewhat smaller degree of contrast (the so-called biasing effect).

\begin{figure*}[th]
\setcaptionmargin{5mm} \onelinecaptionsfalse
\begin{center}
\includegraphics[height=\textwidth,keepaspectratio,angle=270]{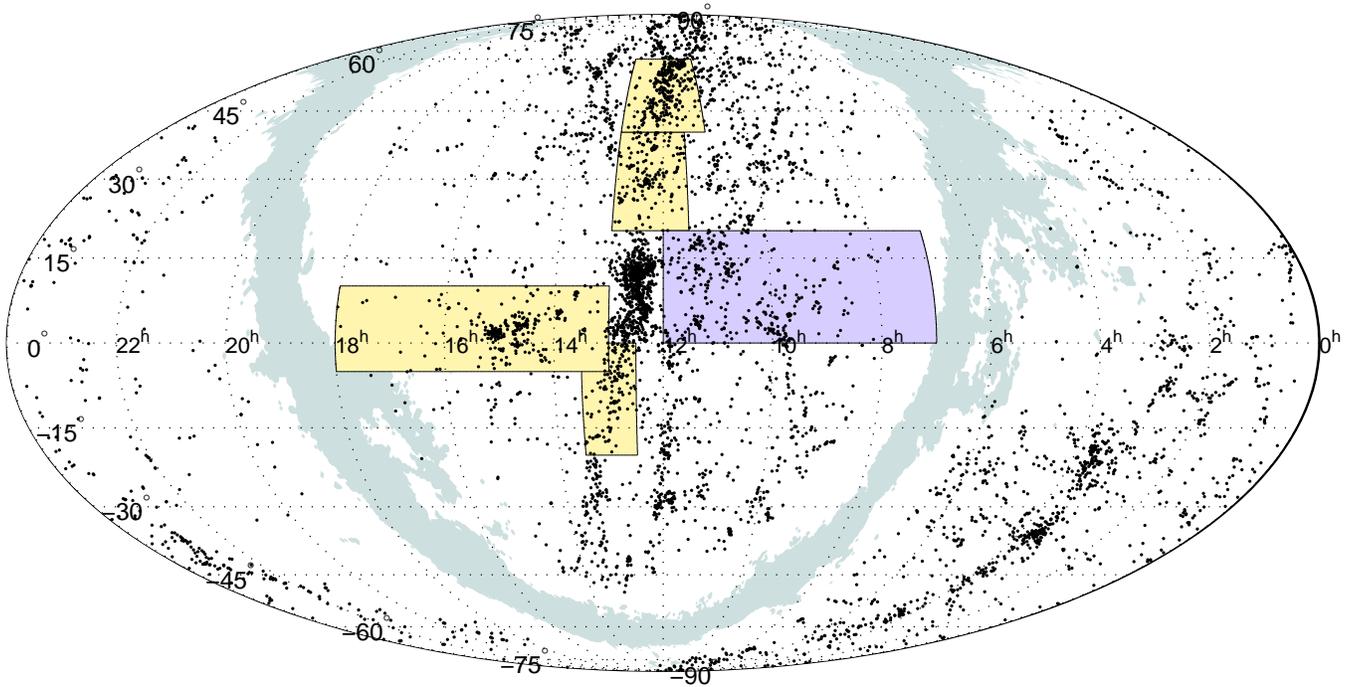}
\end{center}
\caption{\small{}The distribution of galaxies of the
Local Supercluster in equatorial coordinates. The Leo/Can region
and the areas of our previous studies are highlighted in colors.}
\end{figure*}

The analysis of distribution of dark (virial) matter in the most nearby and
well-studied part of the Universe with radial velocities of galaxies of
\mbox{$V_{\rm LG} < 3500$~km\,s$^{-1}$} was conducted by Makarov and
Karachentsev
in~\cite{mak2011:Karachentsev_n,kar2008:Karachentsev_n,mak2009:Karachentsev_n}.
The main and paradoxical result of this research is that the average density of
dark matter in the Local Supercluster and its vicinity
\mbox{$\Omega_m(\mbox{local})=0.08\pm0.02$} proved to be \mbox{3--4}~times
smaller than the global average density,
$\Omega_m(\mbox{global})=0.28$--$0.30$~\cite{spe2007:Karachentsev_n,spe2013:Karachentsev_n}.
The indications of low density of dark matter in the Local Universe have been
already revealed~\cite{ven1984:Karachentsev_n,tul1987:Karachentsev_n}. A survey
of various explanations for the ``missing dark matter'' paradox can be found
in~\cite{kar2012:Karachentsev_n}. One of them is a suggestion that a significant
part of dark matter is stored in the space between the known clusters and groups
of galaxies, in the ``lethargic'' zones, where because of some reasons the
process of star formation did not get triggered. Such dark elements of the
large-scale structure (massive clumps, extended filaments), if they exist
indeed, may manifest themselves both by the effects of weak gravitational
lensing~\mbox{\cite{veg2010:Karachentsev_n,shan2012:Karachentsev_n}} and by
peculiar motions of nearby galaxies~\cite{kar2011:Karachentsev_n}.

To determine the peculiar (non-Hubble) velocity of the galaxy, \mbox{$V_{\rm
pec}=V_{\rm LG}-H_0\,D$}, we have to measure its radial velocity relative to the
centroid of the Local Group, $V_{\rm LG}$, and the distance $D$, adopting a
fixed value of the Hubble parameter $H_0$. The field of peculiar velocities can
be examined in most detail in the closest volumes, where the amount and quality
of data on the galaxy distances is much higher than that in the distant volumes
of space. In the series of previous studies we have reviewed the data on the
velocities and distances of galaxies in three areas located along the equator of
the Local Supercluster: the region Coma\,I~\cite{kar2011:Karachentsev_n} with
\mbox{$V_{\rm LG}<3000$~km\,s$^{-1}$} and coordinates \mbox{$\bigl[{\rm RA} =
11^h\hspace{-0.4em}.\,5$--$13^h\hspace{-0.4em}.\,0$}, \mbox{${\rm Dec} =
+20^{\circ}$...$+40^{\circ}\bigr]$}, the region of
Ursa\,Major~\cite{kar2013:Karachentsev_n} with \mbox{$V_{\rm
LG}<1500$~km\,s$^{-1}$} and \mbox{$\bigl[{\rm RA} =
11^h\hspace{-0.4em}.\,0$--$13^h\hspace{-0.4em}.\,0$}, \mbox{${\rm Dec} =
+40^{\circ}$...$+60^{\circ}\bigr]$}, and the region of the Virgo Southern
Extension~\cite{kar_nas2013:Karachentsev_n} with \mbox{$V_{\rm
LG}<2000$~km\,s$^{-1}$} and \mbox{$\bigl[{\rm RA} =
12^h\hspace{-0.4em}.\,5$--$13^h\hspace{-0.4em}.\,5$}, \mbox{${\rm Dec}=
-20^{\circ}$--$0^{\circ}\bigr]$} as well as the
Bo\"otes~\cite{kar_nas2014:Karachentsev_n} zone with \mbox{$V_{\rm
LG}<2000$~km\,s$^{-1}$} and \mbox{$\bigl[{\rm RA} =
13^h\hspace{-0.4em}.\,0$--$18^h\hspace{-0.4em}.\,0$}, \mbox{${\rm Dec} =
-5^{\circ}$...$+10^{\circ}\bigr]$}, extending perpendicular to the equator of
the Local Supercluster. In the three areas considered, the estimates of virial
average density of matter were in the range of \mbox{$\Omega_m(\mbox{local}) =
0.08$--$0.11$}, and in the Coma\,I region the existence of a dark attractor with
a mass of about \mbox{$2\times 10^{14}\,M_{\odot}$} at a distance of about
$15$~Mpc was suspected.

The distribution of galaxies of the Local Supercluster with radial velocities
$V_{\rm LG}<2000$~km\,s$^{-1}$ is presented in the equatorial coordinates in
Fig.~1. The zone of strong extinction in the Milky Way (the Zone of Avoidance)
is demonstrated by a patchy gray stripe. The regions we have previously
studied---Coma\,I, Ursa\,Major, Virgo\,SE, Bo\"otes, and the new area of our
interest in the constellations of Leo and Cancer with the coordinates
$\bigl[{\rm RA} = 7^h\hspace{-0.4em}.\,0$--$12^h\hspace{-0.4em}.\,0$, ${\rm Dec}
= 0^{\circ}$...$+20^{\circ}\bigr]$---are marked by dark rectangles.

\section{Observational data for the~Leo/Cancer~sample}

\begin{figure}[b]
\setcaptionmargin{5mm} \onelinecaptionsfalse
\includegraphics[height=0.49\textwidth,keepaspectratio,angle=270]{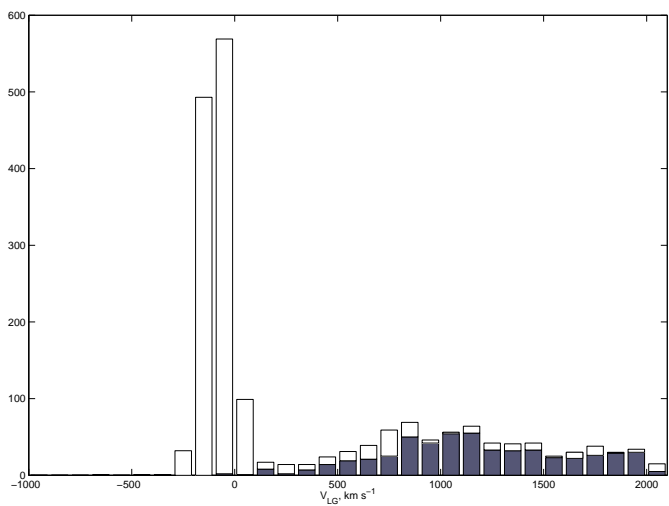}
\caption{\small{}The distribution of 1918 Leo/Can
objects from HyperLeda by radial velocity in the Local Group
rest frame. The true 543 galaxies are shown in dark.}
\end{figure}

\begin{figure*}[ht]
\begin{minipage}[t]{0.49\textwidth}
\setcaptionmargin{5mm} \onelinecaptionsfalse
\includegraphics[height=\textwidth,keepaspectratio,angle=270]{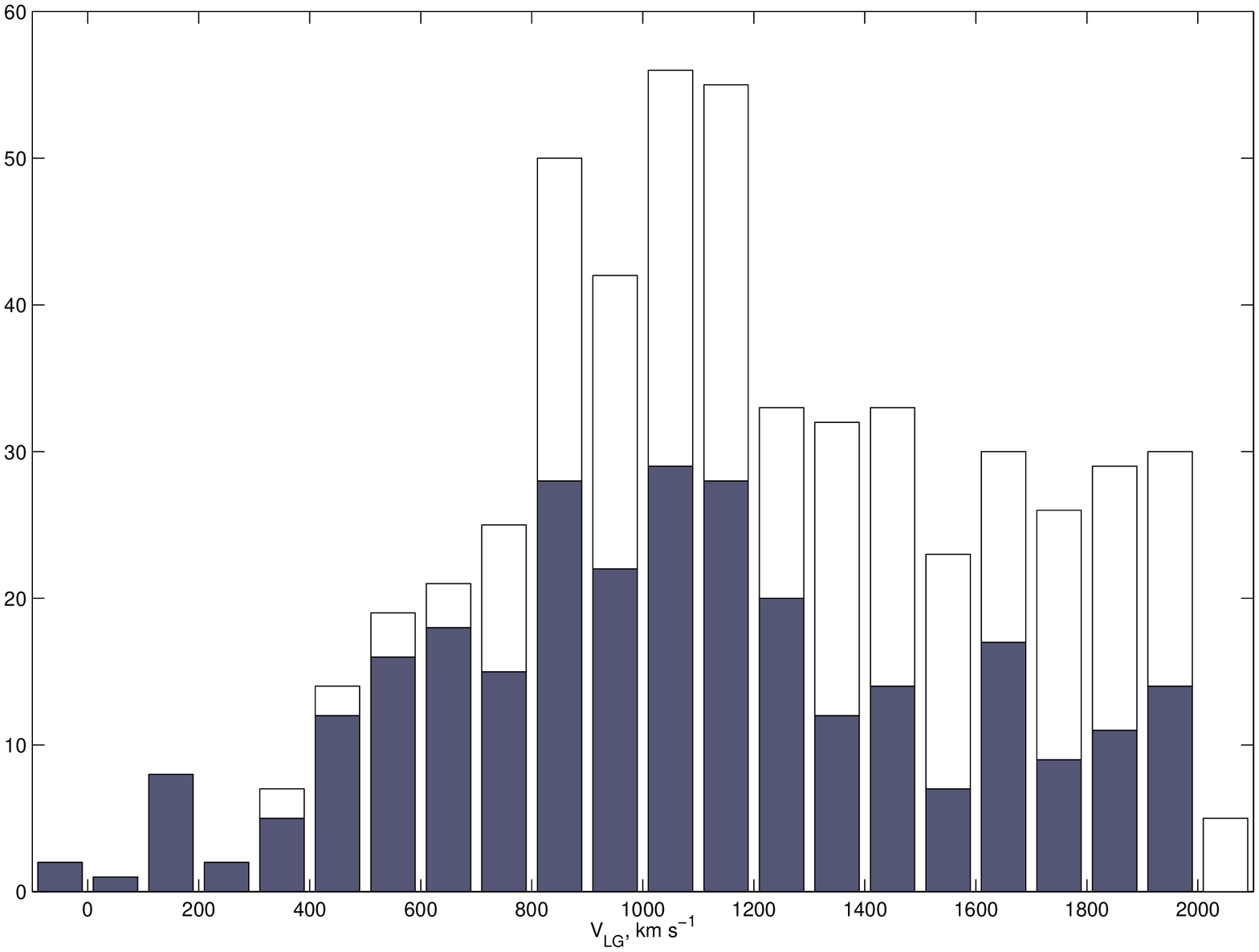}
\caption{\small{}The distribution of 543 galaxies in the
Leo/Can stripe by radial velocity. The 290 objects with measured distances
are marked in dark.}
\end{minipage}
\hfill
\begin{minipage}[t]{0.49\textwidth}
\setcaptionmargin{5mm} \onelinecaptionsfalse
\includegraphics[height=\textwidth,keepaspectratio,angle=270]{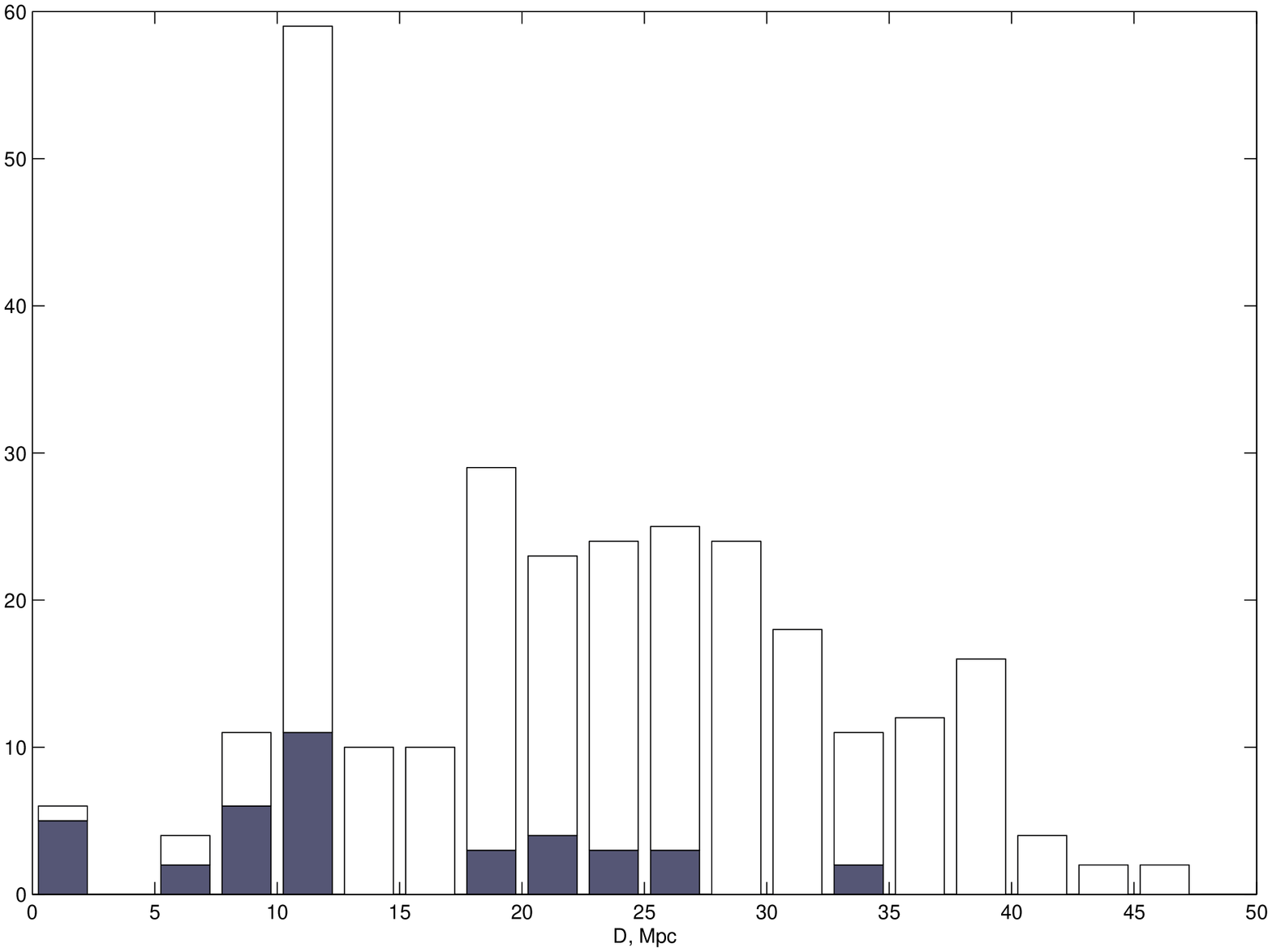}
\caption{\small{}The distribution of 290 galaxies in the
 Leo/Can region by distance. Thirty-nine galaxies with
high-accuracy distance estimates are marked in dark colors.}
\end{minipage}
\end{figure*}

The studied area $20^{\circ}$ in width extends from the virial border of the
Virgo Cluster to the zone of the Milky Way, where the South Pole of the Local
Supercluster is located. A substantial part of the area is covered by the SDSS
optical sky survey~\cite{aba2009:Karachentsev_n}. About 40\% of the Leo/Can
region is covered by the \mbox{ALFALFA} sky survey in the 21~cm line conducted
at the Arecibo radio telescope~\cite{hay2011:Karachentsev_n}. The Leo/Can belt
in its entirety lies in the northern region of the H\,I Parkes All-Sky Survey
(HIPASS) carried out at the Parkes radio
telescope~\cite{wong2006:Karachentsev_n}. The abundance of data on radial
velocities of galaxies, their photometry and H\,I line width made it possible to
determine the distances to galaxies by the Tully--Fisher
relation~\cite{tul1977:Karachentsev_n} and get the local field of peculiar
velocities with high density.

According to HyperLeda\footnote{\tt http://leda.univ-lyon1.fr}, the region
of\linebreak \mbox{$\bigl[{\rm RA} =
7^h\hspace{-0.4em}.\,0$--$12^h\hspace{-0.4em}.\,0$}, \mbox{${\rm Dec} =
0^{\circ}$...$+20^{\circ}\bigr]$} contains\linebreak 1918~objects with radial
velocities in the Local Group rest frame of \mbox{$V_{\rm
LG}<2000$~km\,s$^{-1}$}. Their distribution according to the radial velocity is
shown in Fig.~2. Most of the objects have near zero velocities, being the stars
of our Galaxy. Our analysis of the HyperLeda data has shown that only 543 out of
1918 objects proved to be real galaxies. They are marked in Fig.~2 in dark. The
sample of objects with \mbox{$V_{\rm LG}<2000$~km\,s$^{-1}$} not only contains
the stellar background objects but also a large number of different fragments of
galaxies taken for individual objects in the SDSS. A large proportion of our
initial list was also composed of the so-called ``ghosts,\!'' the dummy
\mbox{H\,I-ALFALFA} survey sources with a low signal-to-noise ratio, not
identified with galaxies. Note that the NED database ({\tt
http://ned.ipac.caltech.edu}) contains yet more than 1000 fictitious
``galaxies'' with $V_{\rm LG}<2000$~km\,s$^{-1}$ in the considered region. All
that indicates that the automatic use of the HyperLeda and NED data without
their careful visual analysis can lead to serious distortions of the researched
field of peculiar velocities of galaxies.

Selecting the galaxies in our list, we also checked and updated their different
characteristics. The summary of the data we used is shown in
Table~1.\footnote{The electronic version of the table is available from
the VizieR database: {\tt
http://cdsarc.u-strasbg.fr/}\linebreak{\tt/viz-bin/qcat?J/other/AstBu/70.1}} The
columns contain: (1)~the name of the galaxy or its number in the known catalogs;
(2)~equatorial coordinates for the epoch J\,2000.0; (3)~radial velocity relative
to the centroid of the Local Group with the parameters of the apex, used in NED;
(4)~the velocity of the galaxy relative to the three-degree CMB radiation with
the parameters of the apex from NED; (5)~integral apparent magnitude of the
galaxy in the $B$-band according to NED or HyperLeda; in the presence of strong
differences in the $B$ values, we have resorted to our own visual apparent
magnitude estimates based on the photometry of other galaxies of similar
structure; (6)~the half-width $W_{50}$ of the 21~cm line, measured at 50\% of
the maximum intensity, the main sources of data on it were the
ALFALFA~\cite{hay2011:Karachentsev_n,sti2009:Karachentsev_n} and
HIPASS~\cite{wong2006:Karachentsev_n} H\,I-surveys with the addition of data
from later publications~\cite{lee2012:Karachentsev_n}; (7)~distance to the
galaxy in Mpc; (8)~a method, with which the distance was measured: rgb---by the
tip of the red giant branch; cep---by Cepheids; SN---by the supernova
luminosity; sbf---by the surface brightness fluctuations; mem---by an obvious
membership of galaxies in known groups; tf, TF, TFb---by the
\mbox{Tully--Fisher} relation between $W_{50}$ and the luminosity of the galaxy,
where the letters tf mark the median distance estimates adopted from NED, the
capital letters TF mark our $D$ estimates by the relation
from~\cite{tul2009:Karachentsev_n}:
$$
M_B=-7.27(\log W^c_{50}-2.5)-19.99, \eqno(1)
$$
where the width $W^c_{50}$ is corrected for the inclination of the galaxy to the
line of sight; as noted in~\cite{mcg2005:Karachentsev_n}, low luminosity
galaxies rich in gas systematically deviate from relation (1) in need of the
so-called ``baryon correction''; in late-type galaxies (Ir,~Im,~Sm) with the
H\,I-magnitude of \mbox{$m_{21}\lesssim m_B$}, where \mbox{$m_{21} = -2.5\log
F(\mbox{H\,I})+17.4$}, and $F(\mbox{H\,I})$ is the flux in the 21~cm line in
Jy\,km\,s$^{-1}$, the hydrogen mass exceeds the mass of the star, therefore, in
cases when \mbox{$m_{21} < m_B$}, we determined the distances by the relation
$$
\log D = 0.2 (m-M) -5, \eqno(2)
$$
using the value of $m_{21}$ instead of $m_B$; two dozens of such gas-rich
galaxies were marked with TFb; (9)~the morphological type of galaxies we have
determined regardless of the NED and HyperLeda data; (10)~the name of the
brightest galaxy in the group to which a given galaxy belongs according
to~\cite{mak2011:Karachentsev_n,kar2008:Karachentsev_n,mak2009:Karachentsev_n}.

\begin{figure*}
\setcaptionmargin{5mm} \onelinecaptionsfalse
\begin{center}
\includegraphics[height=\textwidth,keepaspectratio,angle=270]{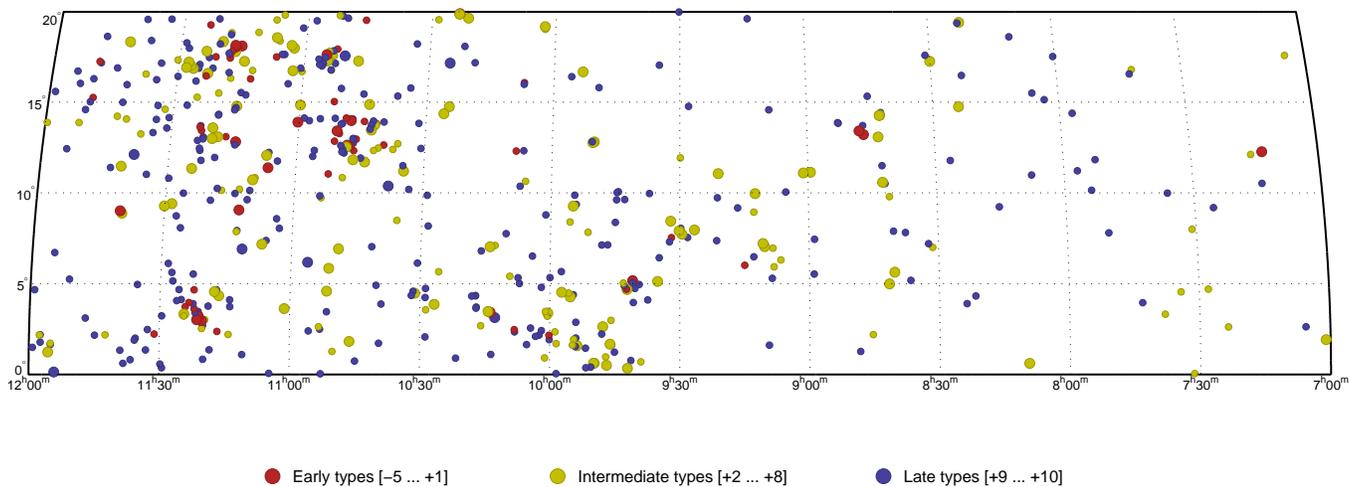}
\end{center}
\caption{\small{}Morphological types of galaxies
in the Leo/Can region. Low-luminosity galaxies, $M_B>-17^{\rm m}$,
are shown by smaller diameter circles.}
\end{figure*}

\begin{figure*}
\setcaptionmargin{5mm} \onelinecaptionsfalse
\begin{center}
\includegraphics[height=\textwidth,keepaspectratio,angle=270]{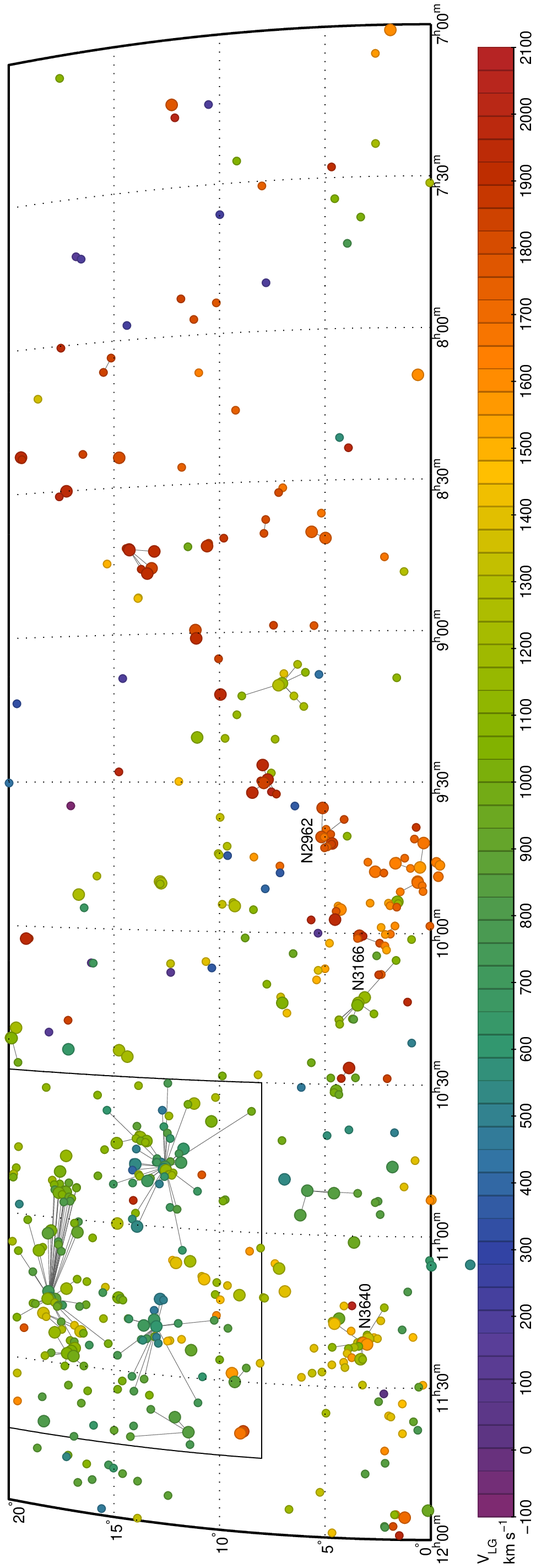}
\includegraphics[height=0.59\textwidth,keepaspectratio,angle=270]{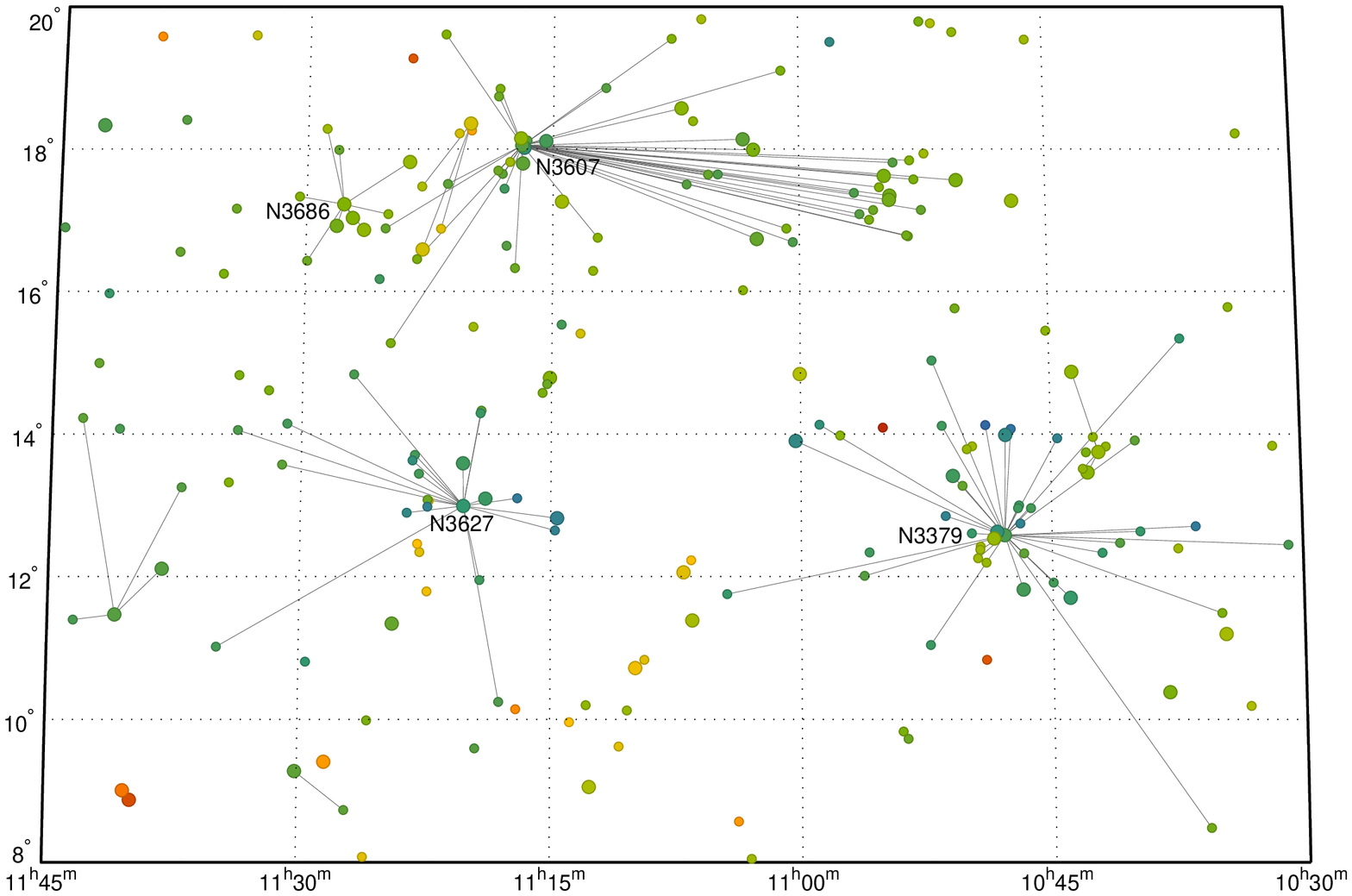}
\end{center}
\caption{\small{}The distribution of galaxies in
Leo/Can by the radial velocity in accordance with the specified
scale (above). The bottom panel represents the region of the
nearby NGC\,3379 and NGC\,3627 groups in close-up. The members of
the group are linked with the main galaxy with lines.}
\end{figure*}

The distribution of 543 galaxies of our sample by the radial velocities $V_{\rm
LG}$ is shown in Fig.~3. Galaxies with individual distance estimates are
marked\linebreak in black. The last interval of the histogram\linebreak
\mbox{$V_{\rm LG}= 2000$--$2100$~km\,s$^{-1}$} captured several galaxies owing
to the differences in the parameters of the apex in NED and HyperLeda. As shown
in Fig.~3, the relative number of galaxies in our sample with known distances
and peculiar velocities is quite large, but their fraction systematically
decreases with increasing radial velocity (distance).

Figure~4 shows the distribution of~290 galaxies in the Leo/Can region by the
distance estimates. Thirty-nine galaxies are marked in black, the distances to
which are measured by the rgb, cep, SN, sbf methods with an error of
approximately $5$--$10$\%. A sharp peak in the histogram falls on the nearby
group members: NGC\,3379~$=$ Leo\,I and NGC\,3627 with the distances of about
$10$--$11$~Mpc. The predominant contribution to the broader secondary maximum at
\mbox{$D= 18$--$32$~Mpc} is given by the groups around the NGC\,2962, NGC\,3166,
NGC\,3227, NGC\,3686, and NGC\,3810 galaxies.

The map of the distribution of galaxies in the Leo/Can stripe by the
morphological types is presented in Fig.~5. The early-type E-S0a galaxies,
spiral Sa--Sm galaxy types and late-type Irr, Im, BCD objects are marked by the
circles of different density. The low luminosity galaxies with \mbox{$M_B>
-17^m\hspace{-0.4em}.\,0$} are illustrated by small circles. According to the
well-known general tendency, late-type dwarf systems are distributed more
uniformly than the galaxies of normal luminosity. Most of the early-type
galaxies are concentrated in groups. However, the isolated E and S0 galaxies are
found even among the field galaxies, generally having low luminosity and
emission features (UGC\,5923, UGC\,6233, IC\,676, IC\,745). The presence in this
region of a compact isolated dE-galaxy \mbox{CGCG\,036-042~$=$} PGC\,02947 has
been the subject of special discussion in~\cite{pau2014:Karachentsev_n}.

The panorama of the distribution of galaxies in our sample by the equatorial
coordinates and radial velocities relative to the centroid of the Local group is
shown in Fig.~6. The top panel of the figure represents the entire the studied
area, indicating the names of the most populated groups, and the bottom panel
presents in a larger scale, the region, occupied by the nearby NGC\,3379 and
NGC\,3627 groups. The radial velocities of galaxies are marked according to the
density scale, shown between the panels. The members of the richest groups are
connected by lines with the corresponding main galaxy of the group. As one can
see, most of the galaxies with radial velocities $V_{\rm LG} <1000$~km\,s$^{-1}$
are located in the left upper corner of the studied area immediately adjacent to
the west border of the Virgo Cluster. Galaxies with the velocities \mbox{$V_{\rm
LG} > 1500$~km\,s$^{-1}$} dominate the right half of the general Leo/Can map.

Figure~7 presents the distribution of galaxies in this area by the distance
according to the scale located under the top panel. The bottom panel shows the
behavior of the running median with the averaging window of
$0^h\hspace{-0.4em}.\,5$. The distribution of galaxies by the distance looks
quite spotty. However, in the area adjacent to the Virgo Cluster (${\rm RA}>
10^h\hspace{-0.4em}.\,4$), the average distance of galaxies of the Leo/Can
stripe approximately corresponds to the distance of the cluster itself of about
$17$~Mpc. The typical distances of galaxies in the median zone \mbox{${\rm RA}=
8^h\hspace{-0.4em}.\,3$--$10^h\hspace{-0.4em}.\,3$} exceed $25$~Mpc, which is
probably due to the presence of a chain of distant groups (NGC\,2648, NGC\,2894,
NGC\,2962, NGC\,3023) crossing this zone diagonally. In the rightmost region of
Fig.~7, the number of galaxies with measured distances is not large, but an
unusual diffuse structure stands out among them. We called it the Gemini Flock.
Seven galaxies it contains are connected in the figure by the common perimeter.
The members of this ``flock'' are dwarf galaxies with active star formation,
they all have anomalously low radial velocities of about $180$~km\,s$^{-1}$ and
rgb distances of around $8$--$9$~Mpc. B.~Tully~\cite{tul2006:Karachentsev_n} was
the first to address this system as an association of four dwarfs. Then three
more members were attributed to
it~\mbox{\cite{kar2006:Karachentsev_n,mcq2014:Karachentsev_n}}, two of which
were recently discovered and studied in the SHIELD
survey~\cite{mcq2014:Karachentsev_n}.

\begin{figure*}
\setcaptionmargin{5mm} \onelinecaptionsfalse
\begin{center}
\includegraphics[height=\textwidth,keepaspectratio,angle=270]{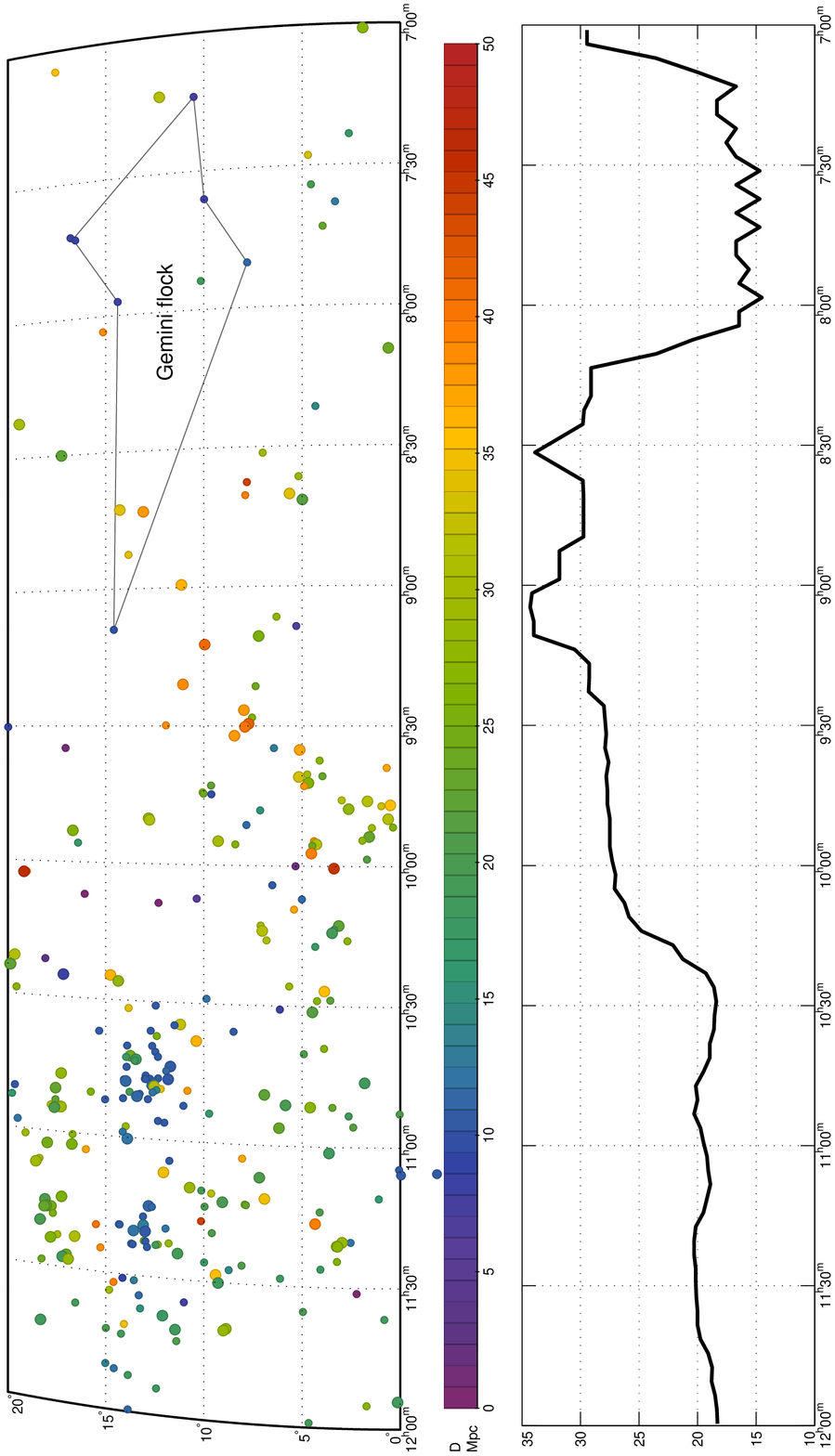}
\end{center}
\caption{\small{}The distribution of galaxies in the Leo/Can by distance
according to the presented scale. The bottom panel
shows the behavior of the median distance along the ${\rm RA}$ with the window of $0^h\hspace{-0.4em}.\,5$.}
\end{figure*}

As follows from the data of Table~1, about 80\% of distance estimates are
inferred by the Tully--Fisher method. The accuracy of the method for normal
luminosity galaxies is about $0^m\hspace{-0.4em}.\,4$, or approximately $20$\%.
When applied to dwarf galaxies, the accuracy of the method decreases because of
uncertainty of the $W_{50}$ correction for the inclination of the galaxy and
other factors. However, when averaged over many members of one group its average
distance can be determined by the \mbox{TF-method} with quite an acceptable
accuracy. Distance estimates via the rgb, cep, SN and sbf methods yield the
accuracy \mbox{2--4}~times better than the TF-method, i.e., one measurement made
by the ``exact'' method is statistically equivalent to about \mbox{$5$--$15$} TF
estimates. However, we encountered cases where the distance estimate made by an
accurate method significantly differed from the \mbox{TF-estimates} for the
other members of the same group. For example, the distance to NGC\,3626 based on
the fluctuations of surface brightness,
$20.0$~Mpc~\cite{ton2001:Karachentsev_n}, looks understated compared with the
average distance of the other members of the group, $26.3$~Mpc. The obvious
reason for underestimating the sbf-distance is caused by the presence in this Sa
galaxy of dust bands that lead to an overestimation of the measured brightness
fluctuations. The other example is the Sc-galaxy NGC\,3389, where the distance
$32.8$~Mpc by SN\,Ia~\cite{par2000:Karachentsev_n} proved to be $10$~Mpc larger
than that found by certain other methods. Both these galaxies are prominent by
large peculiar velocities that are cancelled out by the use of alternative
distance estimates.

\section{Peculiar motions of galaxies in~the~Leo/Cancer~region} The field of
peculiar velocities of galaxies in the stripe considered at the Hubble parameter
of \mbox{$H_0=72$~km\,s$^{-1}$\,Mpc$^{-1}$} is shown in Fig.~8. The upper half
of the figure corresponds to peculiar motions relative to the centroid of the
Local Group, the bottom half---relative to the three-degree microwave radiation.
The marking of peculiar velocities corresponds to the density scale, which in
the first case covers the range of $-2000$~km\,s$^{-1}$ to $+800$~km\,s$^{-1}$,
and in the second case---of $-1400$~km\,s$^{-1}$ to $+1400$~km\,s$^{-1}$. The
broken lines under the panels $V_{\rm pec}$ show the variation of the median
peculiar velocity along the stripe with a window of $0^h\hspace{-0.4em}.\,5$.

\begin{figure*}
\setcaptionmargin{5mm} \onelinecaptionsfalse
\begin{center}
\includegraphics[height=\textwidth,keepaspectratio,angle=270]{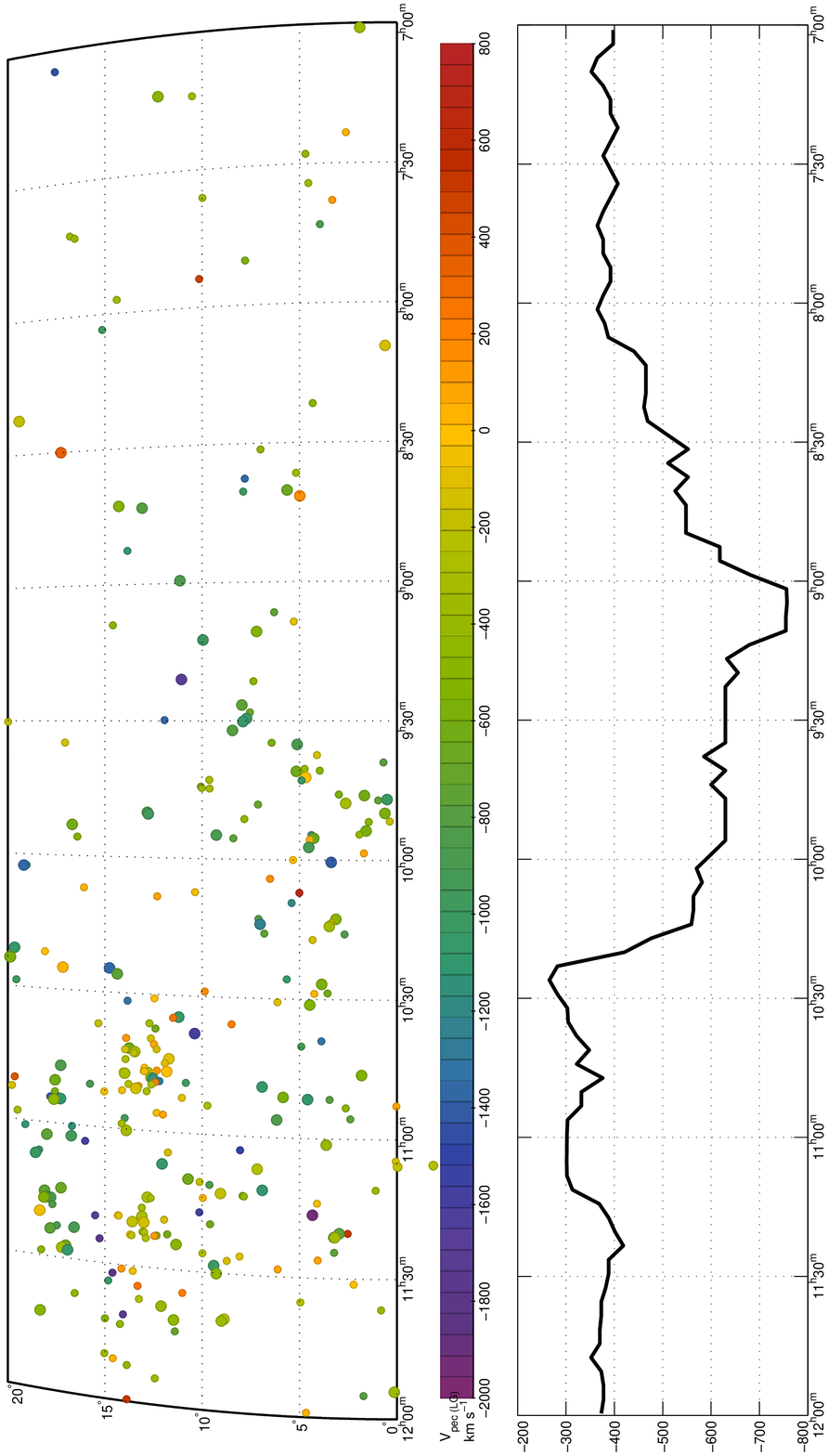}
\includegraphics[height=\textwidth,keepaspectratio,angle=270]{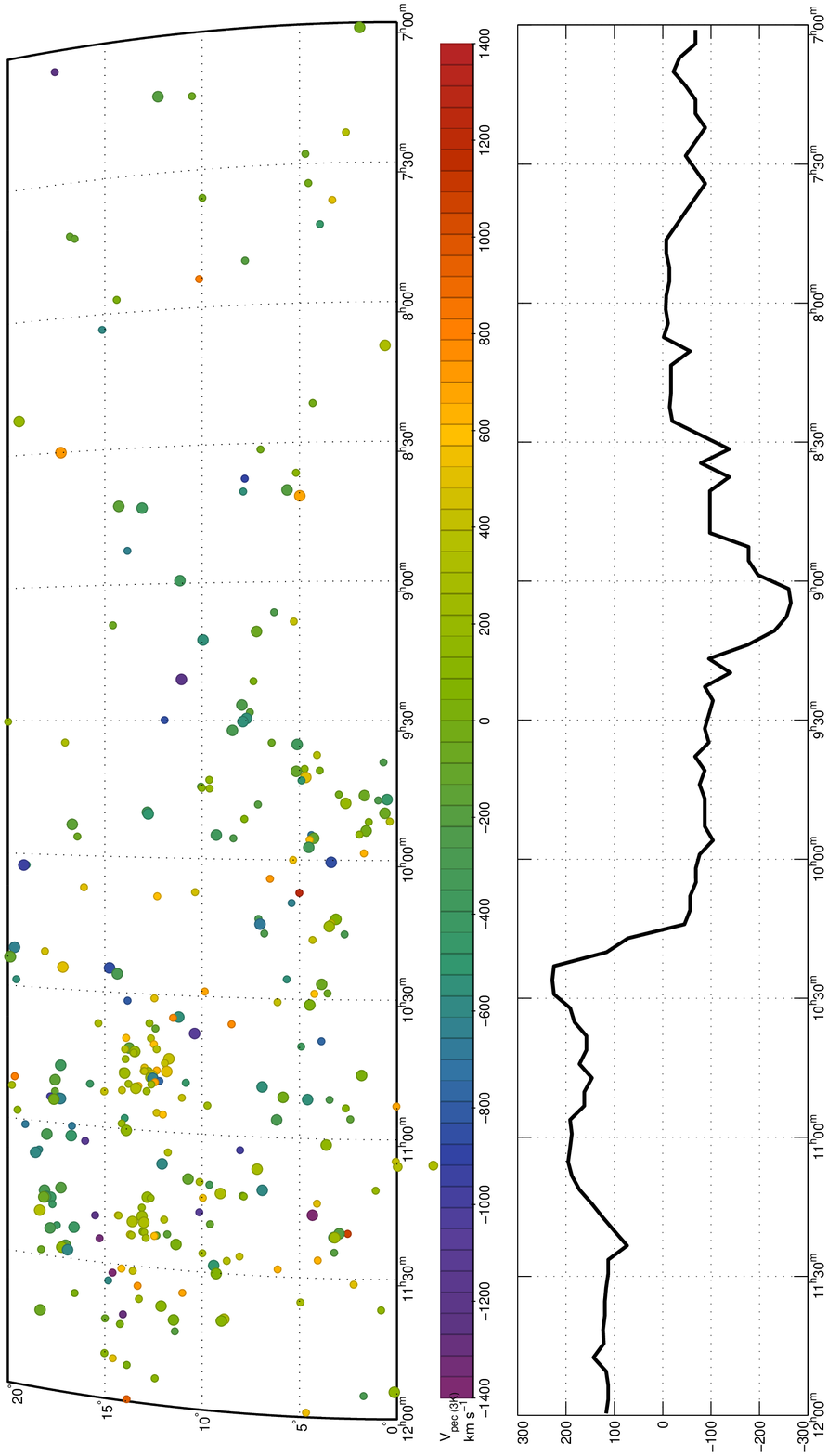}
\end{center}
\caption{\small{}The distribution of galaxies in the Leo/Can by
the scale of peculiar velocities. The top and bottom panels correspond to $V_{\rm pec}$
in the Local Group rest frame and in the three-degree
CMB rest frame. The broken lines show the behavior of the median
peculiar velocity along the stripe.}
\end{figure*}

With an average distance of galaxies of around $25$~Mpc and distance measurement
error by the Tully--Fisher method of about $20$\%, the expected error in the
estimate of peculiar velocity is approximately $360$~km\,s$^{-1}$. The observed
variations of $V_{\rm pec}$ are significantly higher than this value. In the
system of the Local Group the median peculiar velocity remains negative
throughout the ${\rm RA}$ range from the Virgo Cluster to the region of the
Milky Way at high supergalactic latitudes, varying from $-300$~km\,s$^{-1}$ to
$-700$~km\,s$^{-1}$. This fact is known as the local velocity anomaly
phenomenon~\cite{tul1992:Karachentsev_n}, which is explained by the motion of
the Local Group to the Virgo Cluster \mbox{$(12^h\hspace{-0.4em}.\,5,
+12^{\circ})$} at a velocity of approximately $190$~km\,s$^{-1}$ and the
recession from the expanding Local Void in the direction of
($7^h\hspace{-0.4em}.\,0, -3^{\circ})$ at approximately
$260$~km\,s$^{-1}$~\cite{tul2008:Karachentsev_n}. The volume of space and the
number of galaxies involved in this motion remains rather uncertain.

\setcounter{table}{1}
\begin{table*}
\setcaptionmargin{0mm} \onelinecaptionstrue
{\footnotesize{}%
\caption{\small{}Characteristics of galaxy groups}
\medskip
\begin{tabular}{crrccrrcccrcc}
\hhline{=============}
Group & $N_v$& $\langle V_{\rm LG}\rangle$,& $\langle V_{\rm 3K}\rangle$,& $D$, & \multicolumn{1}{c}{$\sigma_v$,} & $R_h$, & $\log M^*$, & $\log M_{\rm H}$,& $\log M_{\rm H}/M^*$& $N_D$& $\langle m-M\rangle$,& $\sigma(m-M)$,\\
& & km\,s$^{-1}$ & km\,s$^{-1}$ & Mpc & km\,s$^{-1}$ & kpc & [$M_{\odot}$] & [$M_{\odot}$]& & & mag & mag \\[+5pt]
(1)&(2)&\multicolumn{1}{c}{(3)} & (4) &(5) & \multicolumn{1}{c}{(6)} & \multicolumn{1}{c}{(7)} & (8) & (9) & (10) &(11)& (12) & (13) \\
\hhline{=============}
NGC\,2648 & 8& 1933 & 2348 &36.0 & 55~~~ & 128 & 11.09 & 11.98& 0.89 & 2~~ & 32.78& 0.10 \\
NGC\,2775 & 9& 1249 & 1740 &26.9 & 89~~~ & 296 & 11.37 & 12.99& 1.62 & 2~~ & 32.15& 0.10 \\
NGC\,2894 & 7& 1952 & 2483 &39.6 & 50~~~ & 458 & 11.32 & 12.23& 0.91 & 4~~ & 32.99& 0.10 \\
NGC\,2962 & 10& 1778 & 2304 &31.6 & 53~~~ & 161 & 10.99 & 11.94& 0.95 & 6~~ & 32.50& 0.32 \\
NGC\,2967 & 6& 1654 & 2262 &35.8 & 62~~~ & 507 & 11.03 & 12.75& 1.72 & 1~~ & 32.77& -- \\
UGC\,5228 & 4& 1683 & 2231 &32.7 & 40~~~ & 188 & 10.31 & 11.90& 1.59 & 2~~ & 32.57& 0.05 \\
NGC\,3023 & 5& 1667 & 2222 &28.8 & 21~~~ & 35 & 10.44 & 11.40& 0.96 & 3~~ & 32.30& 0.17 \\
NGC\,3020 & 3& 1240 & 1723 &30.2 & 45~~~ & 44 & 10.24 & 11.53& 1.29 & 2~~ & 32.40& 0.06 \\
NGC\,3049 & 3& 1297 & 1805 &30.2 & 15~~~ & 144 & 10.29 & 11.31& 1.02 & 1~~ & 32.40& -- \\
UGC\,5376 & 4& 1847 & 2393 &45.3 & 66~~~ & 253 & 10.87 & 12.23& 1.36 & 1~~ & 33.28& -- \\
NGC\,3166 & 10& 1104 & 1742 &20.5 & 44~~~ & 126 & 11.36 & 11.97& 0.61 & 5~~ & 31.56& 0.35 \\
NGC\,3227 & 6& 1054 & 1495 &25.7 & 74~~~ & 128 & 11.27 & 12.50& 1.23 & 5~~ & 32.05& 0.27 \\
NGC\,3338 & 7& 1105 & 1594 &20.1 & 50~~~ & 112 & 10.77 & 11.05& 0.28 & 3~~ & 31.52& 0.33 \\
NGC\,3379 & 36& 702 & 1198 &10.8 & 193~~~ & 191 & 11.53 & 13.10& 1.57 & 14~~ & 30.18& 0.31 \\
NGC\,3423 & 4& 850 & 1389 &23.1 & 21~~~ & 570 & 10.64 & 12.14& 1.50 & 4~~ & 31.82& 0.25 \\
NGC\,3521 & 3& 593 & 1160 &10.7 & 37~~~ & 132 & 11.10 & 12.52& 1.42 & 2~~ & 30.15& 0.00 \\
NGC\,3596 & 3& 1009 & 1483 &14.0 & 42~~~ & 41 & 10.13 & 11.43& 1.30 & 0~~ & 30.73& -- \\
NGC\,3607 & 45& 928 & 1377 &25.0 & 115~~~ & 471 & 11.77 & 13.29& 1.52 & 12~~ & 31.99& 0.28 \\
NGC\,3626 & 5& 1387 & 1833 &25.6 & 86~~~ & 187 & 11.06 & 12.75& 1.69 & 4~~ & 32.04& 0.39 \\
NGC\,3627 & 20& 697 & 1182 &10.8 & 136~~~ & 201 & 11.47 & 12.96& 1.49 & 15~~ & 30.09& 0.38 \\
NGC\,3640 & 14& 1240 & 1785 &27.2 & 134~~~ & 252 & 11.34 & 12.66& 1.32 & 4~~ & 32.17& 0.08 \\
NGC\,3686 & 10& 1057 & 1508 &21.9 & 91~~~ & 175 & 10.97 & 12.65& 1.68 & 5~~ & 31.70& 0.37 \\
NGC\,3810 & 5& 844 & 1328 &17.7 & 43~~~ & 360 & 10.67 & 12.12& 1.45 & 5~~ & 31.24& 0.23 \\
\hhline{-------------}
Mean & 10& 1255 & 1765 &25.7 & 68~~~ & 224 & 10.86 & 12.23& 1.28 & 4~~ & 31.89& 0.22 \\
\hhline{=============}
\end{tabular}}
\end{table*}

In the frame of reference related to the microwave radiation, the median
peculiar velocity varies symmetrically from $+200$~km\,s$^{-1}$ to
\mbox{$-200$~km\,s$^{-1}$}. The positive values of $V_{\rm pec}$(3K) in the area
$10^h\hspace{-0.4em}.\,5$--$12^h\hspace{-0.4em}.\,0$ are mainly due to two rich
nearby groups around NGC\,3379 and NGC\,3627, which are moving away from us to
the Virgo Cluster, revealing a positive velocity component along the line of
sight relative to the observer. Note that the Bo\"otes stripe, which is located
on the other side of Virgo and extends up to the Local Void, also clearly
demonstrates the effect of galaxy infall onto the Virgo
Cluster~\cite{kar_nas2014:Karachentsev_n}.

According to the analysis made in~\cite{tul2008:Karachentsev_n}, the pattern of
motions in the Leo/Can region roughly looks like an approach of two elements of
the local large-scale structure: the Local Volume and the Leo cloud with the
mutual velocity of about $500$~km\,s$^{-1}$. New mass measurements of radial
velocities and distances of galaxies in the Leo/Can confirm the existence of
nearby large-scale flows of galaxies with amplitudes that are comparable to the
virial velocities in rich clusters.

\section{Galaxy systems in the Leo/Cancer~region}

The galaxy clustering algorithm used
in~\cite{mak2011:Karachentsev_n,kar2008:Karachentsev_n,mak2009:Karachentsev_n},
led to the detection in the studied area of 23 groups of galaxies, most of which
have been previously known. Taking into account the new data on the radial
velocities and galaxy distances, the list of these groups is shown in Table~2.
The Table columns contain the following main characteristics of the groups:
(1)~the name of the main galaxy; (2)~the number of members with measured radial
velocities; (3,~4)~the average radial velocity of the group (km\,s$^{-1}$)
relative to the centroid of the Local Group and the three-degree blackbody
radiation; (5)~the distance to the group (Mpc), corresponding to the mean
modulus \mbox{$(m-M)$} of its members; (6)~radial velocity dispersion
(km\,s$^{-1}$); (7)~mean harmonic radius of the group (kpc); (8)~logarithm of
the total stellar mass of the group, estimated from the luminosity of galaxies
in $K$-band assuming $M^* /L_K=1\times M_{\odot}/L_{\odot}$; (9)~logarithm of
the projection (virial) mass of the group, which characterizes the mass of the
group's halo~$M_{\rm H}$, $$M_p=(32/\pi G) (N-3/2)^{-1} \sum_{i=1}^{N} \Delta
V_i^2 R_i, \eqno(3) $$ where $\Delta V_i$ and $R_i$ are the radial velocity and
projection distance of the $i$-th member of the group relative to the center of
the group~\cite{hei1985:Karachentsev_n}, $N$ is the number of members, and $G$
is the gravitational constant; (10)~logarithm of the projection to stellar mass
ratio; (11)~the number of group members with distance estimates; (12,~13)~the
average distance modulus and the mean square scatter of moduli. The last line of
the Table contains the mean values of the presented parameters.

\subsection{NGC\,3379~$=$ Leo\,I and NGC\,3627 groups}
Both groups located at a distance of $10.8$~Mpc are the closest and richest
systems in the Leo/Can region. The recent measurements of distances to the main
galaxies in these groups by rgb: $10.7$~Mpc and
$10.8$~Mpc~\cite{lee2013:Karachentsev_n} are in a remarkable agreement with the
data in Table~2. The general view of both groups is shown in the bottom panel of
Fig.~6. The Leo\,I group contains a significant amount of E, S0, dSph-type
galaxies, indicating its advanced evolutionary status. The literature lists
several estimates of the Leo\,I group virial mass: \mbox{$0.72\times
10^{13}\,M_{\odot}$}~\cite{kar2004:Karachentsev_n}, \mbox{$1.7\times
10^{13}\,M_{\odot}$}~\cite{mak2011:Karachentsev_n} and \mbox{$ 1.7\times
10^{13}\,M_{\odot}$}~\cite{kar2014:Karachentsev_n}, which are in good agreement
with the mass estimate \mbox{$1.26\times 10^{13}\,M_{\odot}$} in Table~2. A
compact group NGC\,3338 with an average radial velocity of \mbox{$V_{\rm LG}=
1105$~km\,s$^{-1}$} and the distance of $20.1$~Mpc is projected at the
north-western outskirts of Leo\,I. An association of members of this more
distant group with the Leo\,I members would bring an asymmetry in the velocity
profile of the Leo\,I group and overestimate the mass of its halo. Another
feature of the Leo\,I group is the presence of a hydrogen ring with the diameter
of about $200$~kpc at it center~\cite{sch1989:Karachentsev_n}. Being projected
on dwarf \mbox{dSph--members} of the group, H\,I-clouds lead to fictitious
radial velocities of the dwarfs. The neighboring group NGC\,3627 has a slightly
lower halo mass and a smaller percentage of the early-type galaxies. A notable
feature of this group is a dwarf galaxy of extremely low surface brightness
AGC\,215414, where more than 95\% of baryons are contained not in the stellar
component but rather in its gaseous
component~\mbox{\cite{kar_mak2008:Karachentsev_n,nik2014:Karachentsev_n}}. The
radius of the ``zero velocity sphere'' for the NGC\,3379 and NGC\,3627 groups is
\mbox{$R_0\simeq1.8$~Mpc}, which is higher than the projection distance between
the group centers. We can conclude from this that both groups will eventually
merge into a single dynamic system.

\subsection{NGC\,3607 group}
According to Tully~\cite{tul1988:Karachentsev_n}, this group, along with the
NGC\,3686 group and other more northern groups, is a member of a scattered Leo
cloud association number 21-1. As we can see from the bottom panel of Fig.~6,
there is a subgroup of galaxies (NGC\,3454/55/57) on the western side of this
group, which is probably in the process of merging with the main body of the
group. By its luminosity and virial mass, the NGC\,3607 group is the most
significant object in the Leo/Can region.

\subsection{Other groups}
What is noteworthy, some groups of galaxies classified
in~\cite{mak2011:Karachentsev_n} as dynamically isolated are in fact associated
with each other, forming hierarchical higher-level structures. For example, some
groups of galaxies around NGC\,2962, NGC\,2967, UGC\,5228, and NGC\,3023 have
similar radial velocities and distance estimates. All these four groups are also
associated with the NGC\,2974 group, which is located outside the southern
boundary of our area. For obvious reasons, the dynamic analysis of such
hierarchical structures is facing difficulties.

\subsection{Gemini Flock}
In the Leo/Can region there are 13 galaxies with radial velocities $V_{\rm
LG}<300$~km\,s$^{-1}$. In addition to the four members of the Local Group
(Leo-T, \mbox{Segue-1}, Leo-I, Leo\,V) and its two neighboring dwarfs (Sex\,B,
Leo\,P), the remaining 7 objects with such velocities are concentrated in the
small area of the sky that occupies 1/10 of the studied area. The probability of
such an event is approximately $10^{-6}$. Taken the nonrandom nature of this
configuration, we obtain a very strange, ephemeral system containing dwarf
galaxies only. With the mean radial velocity \mbox{$\langle V_{\rm LG}\rangle =
190$~km\,s$^{-1}$} the average distance to this group is $8.5$~Mpc, hence, it is
approaching the Local Group at a peculiar velocity of $-423$~km\,s$^{-1}$. This
``flock'' of galaxies in the Gemini constellation has a radial velocity
dispersion $\sigma_v$ of only about $20$~km\,s$^{-1}$ and the projection radius
of about $5^{\circ}$, or $740$~kpc. The virial mass of the group of
\mbox{$M_{\rm vir} \sim 3\times 10^{11}\,M_{\odot}$} corresponds to these
parameters. At the total stellar mass of the group of \mbox{$M^*=0.96\times
10^9\,M_{\odot}$} the virial-to-stellar mass ratio reaches \mbox{$M_{\rm
vir}/M^* \simeq 300$}, i.e., the average density of its dark matter is close to
\mbox{$\Omega_m \simeq 1$}. However, the obtained values should be rather
considered as formal, since the crossing time for such a loose system exceeds
the age of the Universe by 2.5~times, meaning that the members of the Gemini
Flock cannot be linked with each other in a causal way. An exception is a close
pair UGC\,3974 and KK\,65 with the radial velocity difference of
$10$~km\,s$^{-1}$ and the projection distance of the components of $38$~kpc.
Note that in the transition to the reference system of the cosmic microwave
radiation, the radial velocity dispersion of the group members increases up to
\mbox{$\sigma_v = 55$~km\,s$^{-1}$}, while the average peculiar velocity drops
to \mbox{$V_{\rm pec}(\mbox{3K})=-73$~km\,s$^{-1}$} (i.e., the system is
practically at rest with respect to the CMB).

\renewcommand{\topfraction}{1}
\renewcommand{\bottomfraction}{1}
\renewcommand{\baselinestretch}{0.85}
\begin{table*}[htb]
\begin{center}
\setcaptionmargin{0mm} \onelinecaptionstrue \captionstyle{normal}
{\footnotesize{}%
\caption{\small{}Characteristics of pairs of galaxies}
\medskip
\begin{tabular}{lrrrrrrrrcc}
\hhline{===========}
\multicolumn{1}{c}{Name} & \multicolumn{1}{c}{$\langle V_{\rm LG}\rangle$,} &$\langle V_{3K}\rangle$,& \multicolumn{1}{c}{$\Delta V_{12}$,} & \multicolumn{1}{c}{$D$,} & $R_{12}$, & $\log M^*$, & $\log M_{\rm H}$, & $\log M_{\rm H}/M^*$& $N_D$ &$\sigma(m-M)$,\\
& km\,s$^{-1}$ & km\,s$^{-1}$ & km\,s$^{-1}$ & Mpc & \multicolumn{1}{c}{kpc} & \multicolumn{1}{c}{[$M_{\odot}$]} & \multicolumn{1}{c}{[$M_{\odot}$]} & & & mag \\[+5pt]
\multicolumn{1}{c}{(1)} & \multicolumn{1}{c}{(2)} & \multicolumn{1}{c}{(3)} & \multicolumn{1}{c}{(4)} & \multicolumn{1}{c}{(5)} & \multicolumn{1}{c}{(6)} & \multicolumn{1}{c}{(7)} & \multicolumn{1}{c}{(8)} & \multicolumn{1}{c}{(9)} & (10)& (11) \\
\hhline{===========}
UGC\,3974 & 165~~ & 476~ & 10~~~ & 8.0 & 38 & 8.63~~ & 9.65~~~& 1.02 ~~~~~ & 2 & 0.01 \\
KK\,65 & ~~ & ~ & ~~~ & & & ~~ & ~~~& ~~~~~ & & \\
KK\,67 & 1860~~ & 2216~ & 9~~~ & 39.2 & 540 & 8.92~~ &10.71~~~& 1.79 ~~~~~ & 1 & -- \\
KKH\,43 & ~~ & ~ & ~~~ & & & ~~ & ~~~& ~~~~~ & & \\
IC\,2329 & 1954~~ & 2307~ & 43~~~ & 29.8 & 57 & 9.57~~ &11.09~~~& 1.52 ~~~~~ & 1 & -- \\
P\,1590056 & ~~ & ~ & ~~~ & & & ~~ & ~~~& ~~~~~ & & \\
P\,1331483 & 1832~~ & 2283~ & 3~~~ & 42.0 & 516 & 9.05~~ & 9.74~~~& 0.69 ~~~~~ & 2 & 0.25 \\
A\,182493 & ~~ & ~ & ~~~ & & & ~~ & ~~~& ~~~~~ & & \\
UGC\,04524 & 1776~~ & 2248~ & 3~~~ & 27.0 & 343 &10.05~~ & 9.56~~~& $-$0.49~~~~~~ & 2 & 0.49 \\
NGC\,2644 & ~~ & ~ & ~~~ & & & ~~ & ~~~& ~~~~~ & & \\
A\,193802 & 1308~~ & 1806~ & 10~~~ & 25.5 & 36 & 8.36~~ & 9.63~~~& 1.27 ~~~~~ & 2 & 0.17 \\
SDSS\,0944 & ~~ & ~ & ~~~ & & & ~~ & ~~~& ~~~~~ & & \\
NGC\,3044 & 1142~~ & 1694~ & 86~~~ & 22.8 & 42 &10.44~~ &11.56~~~& 1.12 ~~~~~ & 1 & -- \\
PGC\,135730 & ~~ & ~ & ~~~ & & & ~~ & ~~~& ~~~~~ & & \\
AGC\,192959 & 1623~~ & 2162~ & 31~~~ & 35.0 & 129 &10.61~~ &11.16~~~& 0.55 ~~~~~ & 2 & 0.23 \\
NGC\,3055 & ~~ & ~ & ~~~ & & & ~~ & ~~~& ~~~~~ & & \\
LSBCL\,1-099 & 1599~~ & 2151~ & 21~~~ & 21.2 & 232 & 9.75~~ &11.08~~~& 1.33 ~~~~~ & 1 & -- \\
Ark~227 & ~~ & ~ & ~~~ & & & ~~ & ~~~& ~~~~~ & & \\
UGC\,05401 & 1918~~ & 2360~ & 52~~~ & 43.0 & 110 &10.46~~ &11.55~~~& 1.09 ~~~~~ & 2 & 0.16 \\
UGC\,05403 & ~~ & ~ & ~~~ & & & ~~ & ~~~& ~~~~~ & & \\
UGC\,05633 & 1231~~ & 1714~ & 15~~~ & 31.6 & 271 &10.03~~ &10.86~~~& 0.83 ~~~~~ & 2 & 0.30 \\
UGC\,05646 & ~~ & ~ & ~~~ & & & ~~ & ~~~& ~~~~~ & & \\
NGC\,3246 & 1954~~ & 2504~ & 6~~~ & 31.2 & 350 &10.03~~ &10.17~~~& 0.14 ~~~~~ & 2 & 0.28 \\
VIII\,Zw\,081& ~~ & ~ & ~~~ & & & ~~ & ~~~& ~~~~~ & & \\
UGC\,5708 & 1000~~ & 1547~ & 26~~~ & 21.3 & 53 & 9.79~~ &10.62~~~& 0.83 ~~~~~ & 1 & -- \\
SDSS\,10313 & ~~ & ~ & ~~~ & & & ~~ & ~~~& ~~~~~ & & \\
MGC\,0013223 & 1588~~ & 2158~ & 38~~~ & 20.8 & 24 & 8.78~~ &10.63~~~& 1.85 ~~~~~ & 1 & -- \\
PGC\,032664 & ~~ & ~ & ~~~ & & & ~~ & ~~~& ~~~~~ & & \\
PGC\,135768 & 857~~ & 1414~ & 12~~~ & 18.6 & 48 & 8.57~~ & 9.91~~~& 1.34 ~~~~~ & 1 & -- \\
PGC\,032687 & ~~ & ~ & ~~~ & & & ~~ & ~~~& ~~~~~ & & \\
AGC\,213796 & 1219~~ & 1740~ & 9~~~ & 24.1 & 26 & 9.62~~ & 9.40~~~& $-$0.22~~~~~~ & 2 & 0.13 \\
PGC\,034135 & ~~ & ~ & ~~~ & & & ~~ & ~~~& ~~~~~ & & \\
UGC\,06306 & 1513~~ & 2052~ &132~~~ & 21.0 & 18 &10.15~~ &11.58~~~& 1.43 ~~~~~ & 1 & -- \\
NGC\,3611 & ~~ & ~ & ~~~ & & & ~~ & ~~~& ~~~~~ & & \\
PGC\,034965 & 1419~~ & 1959~ & 6~~~ & 19.4 & 161 & 9.34~~ & 9.84~~~& 0.50 ~~~~~ & -- & -- \\
AGC\,214317 & ~~ & ~ & ~~~ & & & ~~ & ~~~& ~~~~~ & & \\
IC\,2828 & 875~~ & 1381~ & 18~~~ & 16.1 & 255 &10.56~~ &10.99~~~& 0.43 ~~~~~ & 2 & 0.28 \\
NGC\,3705 & ~~ & ~ & ~~~ & & & ~~ & ~~~& ~~~~~ & & \\
PGC\,1218832 & 818~~ & 1345~ & 43~~~ & 11.2 & 9 & 8.28~~ &10.28~~~& 2.00 ~~~~~ & -- & -- \\
PGC\,1218144 & ~~ & ~ & ~~~ & & & ~~ & ~~~& ~~~~~ & & \\
\hhline{-----------}
Mean & 1383~~ & 1876~ & 29~~~ & 25.4 & 163 & 9.55~~ &10.50~~~& 0.95 ~~~~~ & 1.6 & 0.23 \\
\hhline{===========}
\end{tabular}}
\end{center}
\vspace{-4mm}
\end{table*}
\renewcommand{\baselinestretch}{1.1}

\subsection{Pairs of galaxies}

From the list of 509 pairs of galaxies in the Local
Supercluster~\cite{kar2008:Karachentsev_n}, 20 pairs are located in the Leo/Can
region. Their main characteristics are presented in Table~3, the columns of
which contain: (1)~the names of the pair components; (2,~3)~the average radial
velocity relative to the centroid of the Local Group and relative to the CMB
radiation; (4)~the radial velocity difference; (5)~the average distance of the
components; (6)~projection separation between the components; (7,~8)~the total
stellar mass and orbital mass, \mbox{$M_{\rm orb}=(16/\pi G)\,\Delta
V_{12}^2\,R_{12}$}; (9)~the orbital-to-stellar mass ratio; (10)~the number of
components with an individual distance estimate; (11)~the difference in distance
moduli of the pair components. The last row of the Table contains the mean
values of the parameters for the binary systems.

As follows from these data, a typical pair of galaxies has the radial-velocity
difference of the components of approximately $30$~km\,s$^{-1}$, the projection
distance between them is approximately $160$~kpc and the halo mass (orbital
mass) to the stellar mass ratio of about $9$.

In the case of pairs of galaxies, as in the case of groups, most of the distance
estimates were made by the Tully--Fisher method, the error of which is taken as
$\sigma(m-M)\simeq 0^m\hspace{-0.4em}.\,4$. The mean squared distance moduli
difference in the last column of Tables~2 and 3 are $0^m\hspace{-0.4em}.\,25$
and $0^m\hspace{-0.4em}.\,26$ respectively. We can conclude therefore that the
membership of galaxies in groups and pairs, selected using
algorithm~\cite{mak2011:Karachentsev_n,kar2008:Karachentsev_n}, is convincingly
confirmed by the subsequent independent estimates of their distances. The
catalogs of groups, triplets and pairs of galaxies in the Local
Supercluster~\mbox{\cite{mak2011:Karachentsev_n,kar2008:Karachentsev_n,mak2009:Karachentsev_n}}
obviously contain only a small percentage of fictitious members.

\section{Hubble diagram in Leo/Can}

\begin{figure}[th]
\setcaptionmargin{5mm} \onelinecaptionsfalse
\includegraphics[height=0.49\textwidth,keepaspectratio,angle=270]{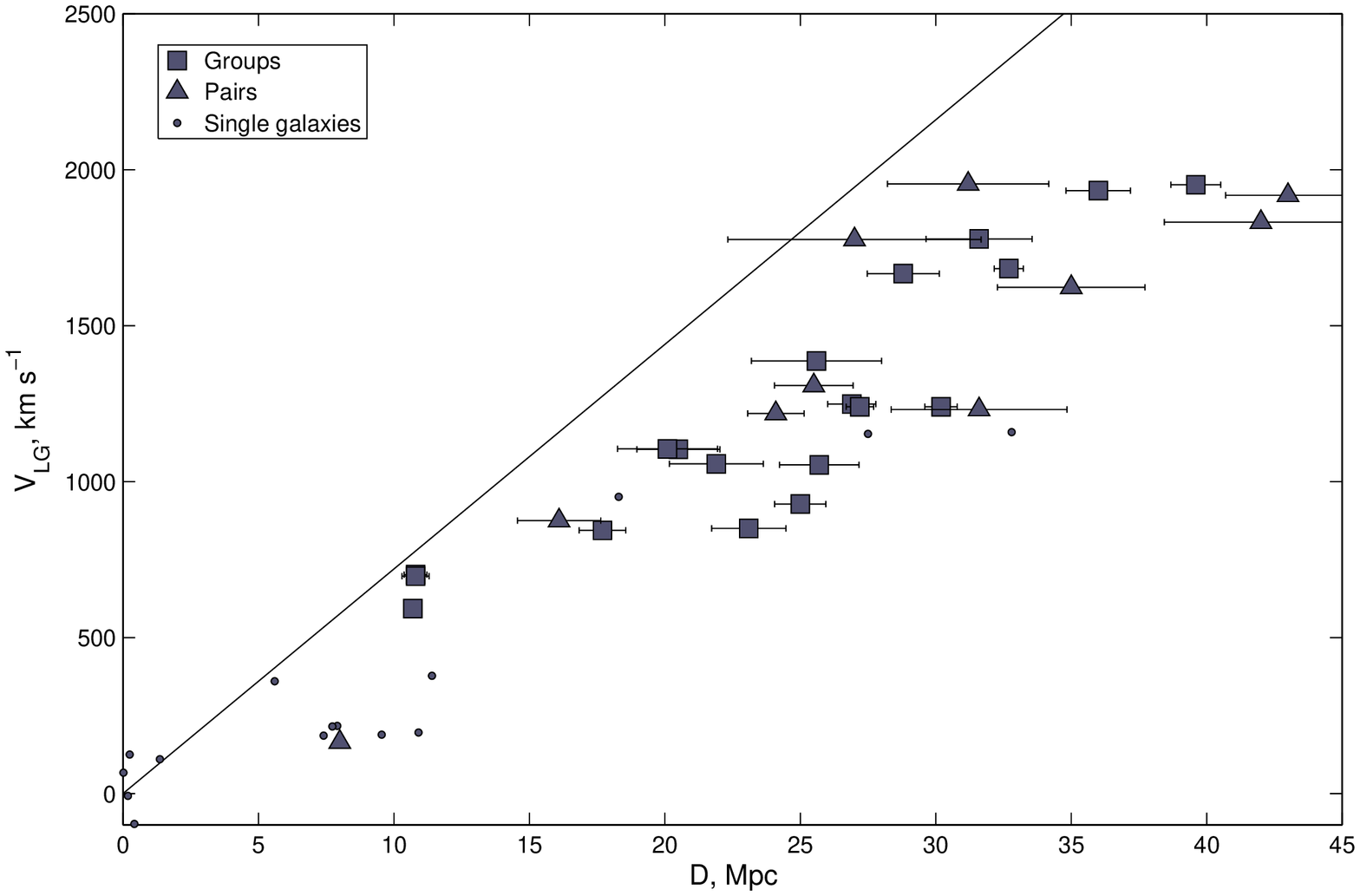}
\vbox{ \vspace{5mm}
\includegraphics[height=0.49\textwidth,keepaspectratio,angle=270]{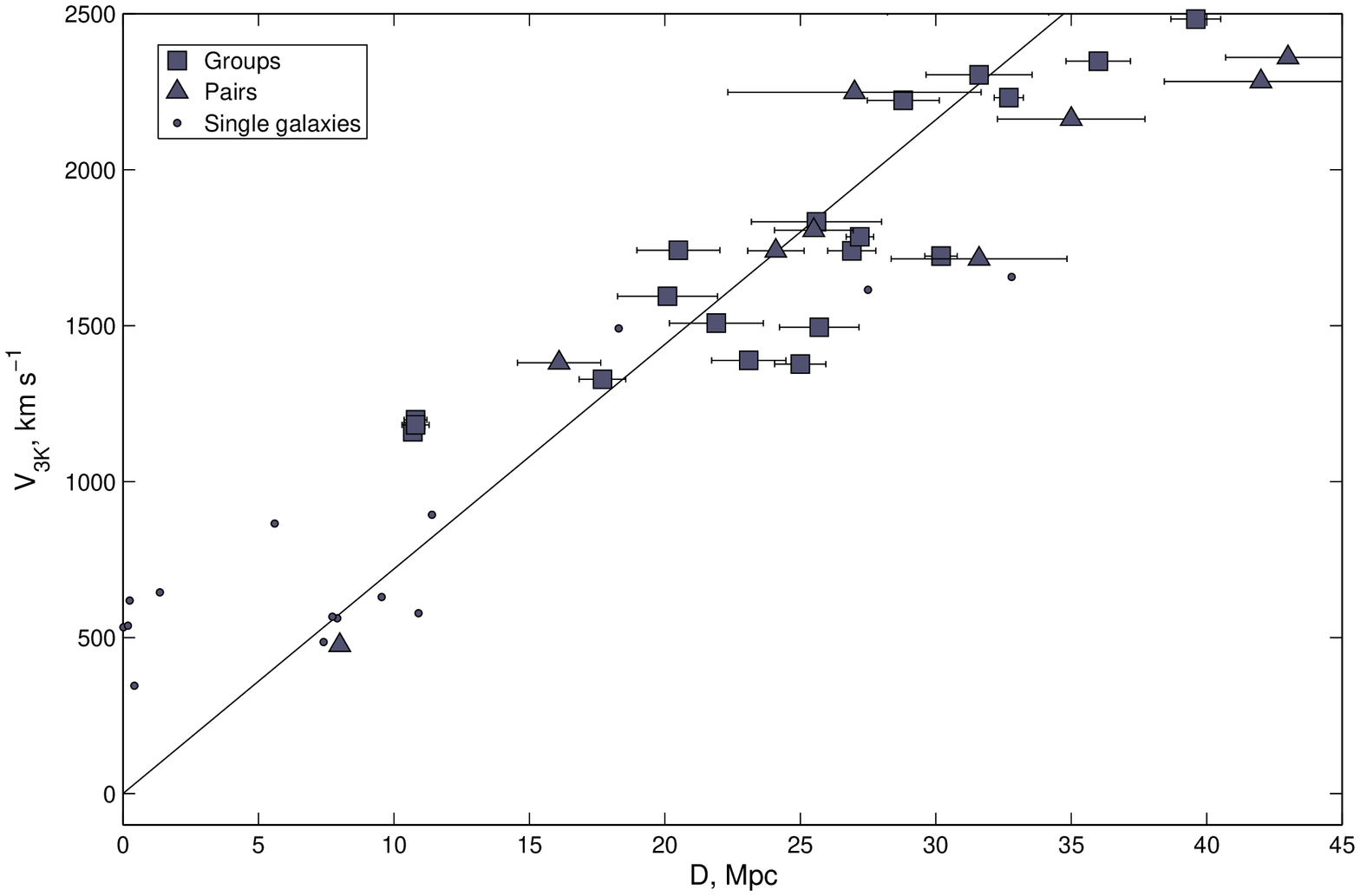}}
\caption{\small{}The Hubble diagram for groups and
pairs of galaxies in the Leo/Can region with respect to the Local
Group system (the top panel) and the CMB (the bottom panel). The
average distance errors are shown by the horizontal segments.}
\end{figure}

The relation between the radial velocities and galaxy distances in the
considered stripe is shown in Fig.~9. The top panel of the figure corresponds to
the velocities in the Local Group rest frame, and the bottom panel depicts the
velocities relative to the cosmic microwave radiation. The squares denote the
groups of galaxies with the number of individual distance estimates $n_D\geq2$,
the triangles---pairs with measured distances for both components, small circles
depict isolated galaxies with high accuracy distance estimates (rgb, cep, SN,
sbf). The straight line on the panels corresponds to the Hubble parameter
\mbox{$H_0=72$~km\,s$^{-1}$\,Mpc$^{-1}$}. We can make the following conclusions
from the presented data.
\begin{list}{}{
\setlength\leftmargin{4mm} \setlength\topsep{2mm}
\setlength\parsep{0mm} \setlength\itemsep{2mm} }

\item[$\bullet$] Almost all the groups and pairs of galaxies have radial
velocities $V_{\rm LG}$ substantially smaller than the ones expected at
\mbox{$H_0=72$~km\,s$^{-1}$\,Mpc$^{-1}$}. The typical velocity shift with
respect to the expected amounts to $\Delta V\sim 500$~km\,s$^{-1}$. In the 3K
system the deviations from the line $H_0=72$~km\,s$^{-1}$\,Mpc$^{-1}$ are not so
great, which indicates the presence in the Local Group of a large peculiar
velocity relative to the CMB.

\item[$\bullet$] Some groups and pairs with well-determined mean distances have
significant peculiar velocities relative to the 3K system. In particular, rich
nearby groups NGC\,3379 and NGC\,3627 as well as a nearby triple system
NGC\,3521 have peculiar velocities of about $+410$~km\,s$^{-1}$, while the rich
group NGC\,3607 has \mbox{$V_{\rm pec} (\mbox{3K})\simeq -420$~km\,s$^{-1}$}.
Large peculiar velocities of these groups are real, they are not caused by the
errors of distances measurements.

\item[$\bullet$] The closest to us diffuse group of dwarf galaxies Gemini Flock
at a high supergalactic latitude is almost at rest in the 3K system. This
distinguishes it from other nearby groups: NGC\,3379, NGC\,3627, and NGC\,3521,
which are located near the Supercluster's equator. As an additional analysis
shows, the volume around the Local Group, which recesses from the Local Void at
a high peculiar velocity, has a flattened shape and is limited by the radius of
about $10$~Mpc.
\end{list}

\section{Local matter density in Leo/Can}

The distribution of 23 groups of galaxies from Table~2 in the considered stripe
based on their virial halo mass estimates and the total stellar mass is shown in
Fig.~10 in the logarithmic scale. In spite of the $M_{\rm H}$ estimate scatter,
primarily caused by the projection factor, there is a positive correlation
between the mass of the dark and luminous matter in groups. It is expressed by
the $\log M_{\rm H} = 1.15\log M^* -0.30$ regression, described in the figure by
the dashed line. The ratio of the sum of orbital masses of binary galaxies from
Table~3 to the sum of their stellar masses, illustrated by the triangle in the
bottom left corner of the figure, also follows the above dependence.

\begin{figure}
\setcaptionmargin{5mm} \onelinecaptionsfalse
\includegraphics[height=0.49\textwidth,keepaspectratio,angle=270]{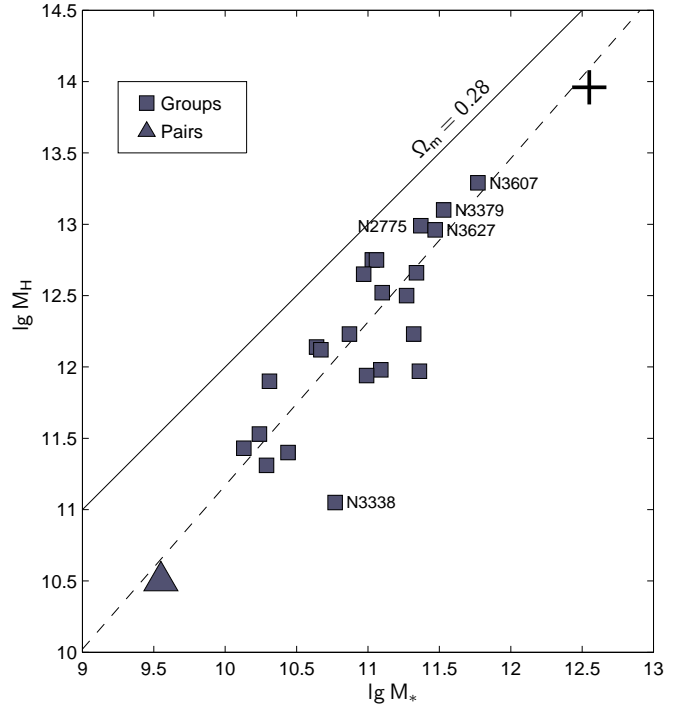}
\caption{\small{}The relationship between the
virial/orbital mass of the halo and the total stellar mass for
groups and pairs of galaxies. The cross shows the ratio
\mbox{$\sum M_{\rm H}/\sum M^* =26$} for the entire volume of
Leo/Can. The solid line corresponds to the global value of $M_{\rm
H}/M^* =97$ at $\Omega_m=0.28$.}
\end{figure}

\onecolumn
\begin{multicols}{2}

The total mass of the halo, contained in the groups and pairs of Leo/Can is
$0.91\times10^{14}\,M_{\odot}$, and their total stellar mass is equal to
$3.5\times10^{12}\,M_{\odot}$. The ratio of these values, $\sum M_{\rm H}/\sum
M^*=26$ is shown in Fig.~10 by a cross. As follows from the data of Table~1,
only 51\% of galaxies in the investigated area are the members of groups and
pairs. However, the field galaxies are predominantly low-luminosity objects. The
computation shows that the additional contribution of single galaxies in the
total stellar mass is only 13\%. They obviously bring some contribution to the
total mass of dark matter too, but with little effect on the ratio of $\sum
M_{\rm H}/\sum M^*$.

According to~\cite{jon2006:Karachentsev_n}, 
matter in the Local Universe is \mbox{$4.5\times 10^8~M_{\odot}$\,Mpc$^{-3}$} at
\mbox{$H_0=72$~km\,s$^{-1}$\,Mpc$^{-1}$}, while the average cosmic density of
matter \mbox{$\Omega_m=0.28$} at \mbox{$H_0=72$} corresponds to the value of
\mbox{$4.5\times 10^{10}~M_{\odot}$\,Mpc$^{-3}$}. The ratio of these global
values, equal to $97$ is shown in Fig.~10 by the solid line. As we can see, all
the systems of galaxies in the Leo/Can are located below the line
\mbox{$\Omega_m=0.28$}. The ratio \mbox{$\sum M_{\rm H}/\sum M^*=26$} for them
corresponds to the local average density of \mbox{$\Omega_m(\rm{local})=0.074$},
which is significantly lower than the global density. This result is in
agreement with the estimate of the mean local density of virial masses we made
in other parts of the structure of the Local
Supercluster~\mbox{\cite{kar2011:Karachentsev_n,kar2013:Karachentsev_n,kar_nas2013:Karachentsev_n,kar2014:Karachentsev_n}}.

Curiously, the line of regression $\log M_{\rm H} =$\linebreak $1.15\log M^*
-0.30$ crosses the line $\Omega_m=0.28$ at the values $\log(M^*
/M_{\odot})\simeq15.3$ and $\log(M_{\rm H} /M_{\odot})\simeq17.3$, which roughly
corresponds to the parameters of the ``homogeneity cell'' with the diameter of
$200$~Mpc. This fact may have a deeper meaning than a mere coincidence.

\section{Final remarks}
The region of the sky in the Leo, Cancer and Gemini constellations, extending
between the center of the Local Supercluster and its South Pole is known as the
``local velocity anomaly.\!'' We have built the map of the distribution of
peculiar velocities of galaxies in it, using the distance estimates for
290~galaxies. We calculated more than a half of distance estimates by the
Tully--Fisher method based on the data on the \mbox{H\,I-line} widths for the
galaxies detected within the HIPASS and ALFALFA H\,I surveys. In the stripe
sized \mbox{$75^{\circ}\times 20^{\circ}$} with a median depth of about
$25$~Mpc, there are 23 groups and 20 pairs of galaxies for which the
virial/orbital mass are determined. In the reference frame related to the
centroid of the Local Group, the majority of groups and pairs have negative
peculiar velocities of about $500$~km\,s$^{-1}$. Relative to the system of the
cosmic microwave radiation, the velocities of most of the distant groups are
small, but the nearby groups NGC\,3379 and NGC\,3627 along with the Local Group
move towards the Leo cloud with the characteristic velocity of
\mbox{$400$--$500$~km\,s$^{-1}$}. Much of this velocity is caused by the motion
of the Local flat ``pancake'' with the diameter of approximately $ 20$~Mpc away
from the center of the Local Void and a fall of the ``pancake'' towards the
Virgo Cluster~\cite{tul2008:Karachentsev_n}.

At a high supergalactic latitude \mbox{${\rm SGB}\simeq-50^{\circ}$} at a
distance of $D\simeq8$~Mpc, an unusual diffuse group called the Gemini Flock was
noted, consisting of seven dwarf galaxies. The characteristic size of the group
of $740$~kpc and the dispersion of radial velocities of $20$~km\,s$^{-1}$ lead
to an estimate of its virial mass of $M_{\rm vir}\simeq 3\times
10^{11}\,M_{\odot}$, which is 300 times larger than the total stellar mass.

The total mass of the halo contained in all\linebreak the groups and pairs of
the Leo/Can region is\linebreak \mbox{$0.9\times10^{14}\,M_{\odot}$}, and the
ratio of this mass to the total stellar mass is $26$. Such a ratio is much lower
than the global value of $M_{\rm H}/M^*=97$ which stems from the standard
cosmological model with the parameter $\Omega_m=0.28$. We conclude from these
data that the problem of ``missing dark matter'' in the Local Universe yet
continues to be an unsolved mystery.

{\bf Acknowledgments}

\noindent{}This work was supported by the RFBR grant\linebreak
\mbox{13-02-90407} and the Foundation for Basic Research of Ukraine, grant
No.~F53.2/15. O.~Nasonova thanks the non-profit Dmitry Zimin's \mbox{\it
Dynasty} foundation and the Russian Science Foundation (grant
\mbox{14-12-00965}) for the financial support.

{}

\end{multicols}

\begin{flushright}
{\it Translated by A.~Zyazeva}
\end{flushright}

\setcounter{table}{0}
\onecolumn
\begin{center}\small{}
\begin{longtable}{lcrrcrrccc}
\caption{The original observational data for 543 galaxies in the Leo/Can region}\\
\hhline{==========}
\multicolumn{1}{c}{Name} & RA (2000.0) Dec & \multicolumn{1}{c}{$V_{\rm LG}$,}& \multicolumn{1}{c}{$V_{\rm 3K}$,}& $B_t$, & \multicolumn{1}{c}{$W_{50}$,} & \multicolumn{1}{c}{$D$,} & Method & Type & Group \\[-5pt]&&\multicolumn{1}{c}{km\,s$^{-1}$}&\multicolumn{1}{c}{km\,s$^{-1}$}&mag&\multicolumn{1}{c}{km\,s$^{-1}$}&\multicolumn{1}{c}{Mpc}&&&\\
\multicolumn{1}{c}{(1)}& (2) & \multicolumn{1}{c}{(3)} & \multicolumn{1}{c}{(4)} & (5)& \multicolumn{1}{c}{(6)} & \multicolumn{1}{c}{(7)}& (8)& (9)& (10) \\
\hhline{==========}\endfirsthead
\hhline{==========}
\multicolumn{1}{c}{(1)}& (2) & \multicolumn{1}{c}{(3)} & \multicolumn{1}{c}{(4)} & (5)& \multicolumn{1}{c}{(6)} & \multicolumn{1}{c}{(7)}& (8)& (9)& (10) \\
\hhline{==========}\endhead
\hhline{----------}\endfoot
\hhline{==========}\endlastfoot
UGC\,03630 & 070103.3+015441 & 1607 & 1929 & 13.98 & 342~~~ & 27.1 & tf & Sb & \\
UGC\,03658 & 070440.0+173457 & 1091 & 1330 & 16.8 & 146~~~ & 35.8 & TFb & Sdm & \\
PGC\,2802325 & 070538.7+023720 & 1590 & 1918 & 18.30 & ~~~ & & & Ir & \\
NGC\,2350 & 071312.2+121558 & 1793 & 2082 & 13.30 & 345~~~ & 31.8 & TF & S0a & \\
UGC\,03755 & 071351.7+103116 & 186 & 486 & 14.10 & ~~~ & 7.41 & rgb & Im & \\
UGC\,03775 & 071552.6+120654 & 2019 & 2314 & 16.4 & ~~~ & & & Sm & \\
UGC\,03830 & 072330.5+023657 & 1232 & 1595 & 14.99 & ~~~ & 16.7 & tf & Scd & \\
PGC\,020981 & 072539.0+091059 & 1064 & 1394 & 16.29 & 94~~~ & & & Ir & \\
AGC\,171494 & 072753.6+044146 & 1928 & 2288 & 18.0 & 120~~~ & 33.0 & TFb & Sd & \\
AGC\,171462 & 073059.7+075935 & 1737 & 2084 & 17.20 & 84~~~ & & & Sm & \\
UGC\,03895 & 073123.4+000312 & 1276 & 1666 & 16.17 & 38~~~ & & & Sm & \\
UGC\,03912 & 073412.6+043247 & 1063 & 1435 & 14.72 & 154~~~ & 20.0 & TF & Sd & \\
AGC\,174585 & 073610.3+095911 & 217 & 562 & 17.9 & 21~~~ & 7.91 & rgb & Ir & \\
UGC\,03946 & 073759.6+031858 & 1026 & 1411 & 14.29 & 90~~~ & 12.7 & TF & Sm & \\
UGC\,3974 & 074155.4+164809 & 160 & 471 & 13.60 & 55~~~ & 8.05 & rgb & Sm & UGC\,3974 \\
KK~65 & 074231.8+163340 & 170 & 484 & 15.30 & ~~~ & 8.02 & rgb & Ir & UGC\,3974 \\
PGC\,021644 & 074308.8+035659 & 768 & 1159 & 15.87 & 91~~~ & 22.8 & TF & Ir & \\
AGC\,174605 & 075021.7+074740 & 196 & 578 & 18.0 & 24~~~ & 10.91 & rgb & Ir & \\
AGC\,174514 & 075230.9+114940 & 1846 & 2207 & 19.0 & ~~~ & & & Ir & \\
AGC\,174616 & 075348.8+100851 & 1796 & 2169 & 17.0 & 55~~~ & 18.4 & TF & BCD & \\
SDSS\,J07565 & 075651.0+111300 & 1831 & 2203 & 18.33 & ~~~ & & & BCD & \\
UGC\,04115 & 075702.1+142328 & 215 & 567 & 15.20 & 76~~~ & 7.73 & rgb & Im & \\
PGC\,1534834 & 080023.9+173127 & 1948 & 2284 & 17.5 & ~~~ & & & BCD & \\
KK~67 & 080324.6+150828 & 1864 & 2220 & 17.4 & 91~~~ & 39.2 & TFb & Ir & KK~67 \\
KKH~43 & 080612.4+153015 & 1855 & 2213 & 19.0 & 49~~~ & & & Ir & KK~67 \\
AGC\,712516 & 080738.8+105850 & 1640 & 2030 & 17.61 & 24~~~ & & & BCD & \\
UGC\,04254 & 080924.0+003634 & 1610 & 2061 & 14.30 & 152~~~ & 24.2 & TF & Sdm & \\
SDSS\,J08103 & 081030.7+183704 & 1383 & 1726 & 18.33 & ~~~ & & & Im & \\
AGC\,182466 & 081532.7+091358 & 1746 & 2158 & 18.7 & 80~~~ & & & Ir & \\
AGC\,188955 & 082137.0+041901 & 575 & 1024 & 17.5 & 38~~~ & 14.5 & TF & Ir & \\
IC\,2329 & 082219.5+192457 & 1975 & 2328 & 15.01 & 180~~~ & 29.8 & TF & Sdm & IC\,2329 \\
SDSS\,J08224 & 082241.4+162851 & 1862 & 2237 & 18.48 & ~~~ & & & Im & \\
PGC\,1590056 & 082246.3+192229 & 1932 & 2286 & 18.3 & ~~~ & & & BCD & IC\,2329 \\
AGC\,188957 & 082341.0+035350 & 1983 & 2437 & 18.6 & ~~~ & & & Ir & \\
UGC\,04385 & 082352.0+144508 & 1832 & 2220 & 14.51 & 148~~~ & & & Sm & \\
AGC\,188875 & 082630.5+114711 & 1740 & 2151 & 17.7 & 27~~~ & & & Ir & \\
UGC\,04444 & 083001.7+171536 & 1956 & 2335 & 14.41 & ~~~ & 22.4 & tf & Scd & \\
PGC\,1536571 & 083104.5+173543 & 2009 & 2387 & 17.9 & ~~~ & & & BCD & \\
PGC\,023907 & 083121.6+070000 & 1677 & 2124 & 16.3 & 126~~~ & 29.7 & TFb & Sm & \\
PGC\,1316080 & 083218.1+071156 & 1836 & 2283 & 18.32 & 39~~~ & & & Ir & \\
SDSS\,J08363 & 083633.4+051041 & 1684 & 2148 & 17.59 & 61~~~ & 28.5 & TF & Ir & \\
PGC\,1331483 & 083735.5+074831 & 1834 & 2285 & 16.63 & 121~~~ & 44.6 & TFb & BCD & PGC\,1331483 \\
UGC\,04524 & 084014.3+053803 & 1757 & 2223 & 15.17 & 158~~~ & 33.9 & tf & Scd & NGC\,2644 \\
AGC\,182493 & 084022.0+075324 & 1831 & 2284 & 17.5 & 94~~~ & 39.4 & TFb & Ir & PGC\,1331483 \\
SDSS\,J08410 & 084105.9+094730 & 1881 & 2324 & 18.06 & ~~~ & & & Sm & \\
NGC\,2644 & 084131.9+045850 & 1754 & 2226 & 13.79 & 213~~~ & 21.6 & TF & Sdm & NGC\,2644 \\
SDSS\,J08420 & 084200.0+103053 & 1848 & 2287 & 16.86 & ~~~ & & & Ir & \\
SDSS\,J08422 & 084226.9+142533 & 1905 & 2319 & 17.80 & ~~~ & & & BCD & NGC\,2648 \\
SDSS\,J08423 & 084232.6+141718 & 2046 & 2461 & 18.15 & ~~~ & & & Im & NGC\,2648 \\
UGC\,04540 & 084235.0+103505 & 1884 & 2323 & 13.8 & 139~~~ & & & Sdm & \\
SDSS\,J08423 & 084235.7+112950 & 976 & 1410 & 18.77 & ~~~ & & & Ir & \\
NGC\,2648 & 084239.8+141707 & 1917 & 2332 & 12.77 & 390~~~ & 34.3 & tf & Sb & NGC\,2648 \\
PGC\,024469 & 084248.2+141555 & 1973 & 2388 & 15.14 & 244~~~ & & & Smp & NGC\,2648 \\
UGC\,04550 & 084315.9+130509 & 1919 & 2343 & 14.74 & 276~~~ & 37.7 & tf & Sb & NGC\,2648 \\
PGC\,1218919 & 084522.5+021124 & 1687 & 2178 & 16.99 & ~~~ & & & Sm & \\
SDSS\,J08452 & 084525.4+151946 & 1504 & 1915 & 18.61 & ~~~ & & & Ir & \\
UGC\,04590 & 084640.0+131249 & 1839 & 2266 & 15.00 & ~~~ & & & S0a & NGC\,2648 \\
PGC\,024666 & 084647.3+134224 & 1941 & 2365 & 16.33 & 72~~~ & & & BCD & NGC\,2648 \\
UGC\,04599 & 084741.7+132509 & 1924 & 2351 & 14.98 & 148~~~ & & & S0 & NGC\,2648 \\
PGC\,1189545 & 084818.7+011551 & 1294 & 1794 & 17.20 & ~~~ & & & Im & \\
SDSS\,J08523 & 085233.8+135028 & 1365 & 1794 & 17.30 & 76~~~ & 34.0 & TF & Im & \\
SDSS\,J08524 & 085240.9+135157 & 1430 & 1859 & 19.70 & ~~~ & & & Ir & \\
SDSS\,J08584 & 085843.9+072631 & 1840 & 2317 & 17.57 & 44~~~ & & & Ir & \\
AGC\,182489 & 085854.1+053134 & 1791 & 2279 & 18.9 & 98~~~ & & & Ir & \\
UGC\,04712 & 085923.6+110806 & 1827 & 2282 & 15.73 & 149~~~ & 36.8 & TF & Scd & \\
NGC\,2725 & 090103.2+110553 & 1913 & 2370 & 14.37 & 186~~~ & & & Scd & \\
PGC\,1373747 & 090515.7+100234 & 1870 & 2337 & 18.36 & 85~~~ & & & BCD & \\
UGC\,04781 & 090634.4+061813 & 1259 & 1751 & 15.40 & 146~~~ & 28.0 & tf & Sd & NGC\,2775 \\
UGC\,04797 & 090810.6+055539 & 1122 & 1617 & 14.58 & 73~~~ & & & Sm & NGC\,2775 \\
AGC\,192558 & 090824.0+065705 & 1390 & 1880 & 17.14 & 35~~~ & & & Sdm & NGC\,2775 \\
KKH~46 & 090836.5+051729 & 409 & 908 & 17.1 & 27~~~ & 6.7 & TF & Ir & \\
LSBC\,D634-0 & 090853.7+143459 & 189 & 630 & 17.5 & 47~~~ & 9.55 & rgb & Ir & \\
PGC\,1200328 & 090920.1+013651 & 1110 & 1630 & 17.30 & ~~~ & & & BCD & \\
NGC\,2775 & 091020.1+070217 & 1169 & 1660 & 11.14 & 409~~~ & & & Sa & NGC\,2775 \\
PGC\,213577 & 091028.7+071118 & 1342 & 1832 & 17.0 & ~~~ & & & Im & NGC\,2775 \\
NGC\,2777 & 091041.8+071224 & 1307 & 1797 & 14.31 & 107~~~ & 25.7 & TF & Sm & NGC\,2775 \\
UGC\,04845 & 091225.8+095720 & 1947 & 2422 & 15.25 & 230~~~ & 41.3 & tf & Scd & \\
SDSS\,J09124 & 091246.6+085620 & 1127 & 1609 & 18.3 & 79~~~ & & & Sm & NGC\,2775 \\
SDSS\,J09125 & 091250.9+062833 & 1285 & 1782 & 17.79 & ~~~ & & & Ir & NGC\,2775 \\
PGC\,1599237 & 091339.0+193708 & 324 & 732 & 17.4 & ~~~ & 8.9 & mem & Im & NGC\,2903 \\
SDSS\,J09145 & 091457.3+060019 & 1242 & 1743 & 17.55 & ~~~ & & & dE & NGC\,2775 \\
AGC\,198354 & 091630.9+091024 & 1162 & 1645 & 18.8 & 27~~~ & & & Ir & \\
AGC\,190238 & 092059.6+110333 & 1116 & 1591 & 15.82 & 142~~~ & 39.1 & TF & Sm & \\
SDSS\,J09211 & 092115.0+094352 & 1199 & 1683 & 18.01 & 23~~~ & & & BCD & \\
AGC\,193816 & 092127.2+072152 & 1208 & 1707 & 17.5 & 56~~~ & 23.6 & TF & Ir & \\
NGC\,2882 & 092636.1+075715 & 1974 & 2473 & 13.51 & 284~~~ & 37.9 & tf & Sc & NGC\,2894 \\
PGC\,1466669 & 092755.6+144559 & 1980 & 2435 & 17.64 & ~~~ & & & Im & \\
AGC\,198454 & 092811.3+073237 & 1193 & 1696 & 18.4 & 45~~~ & 26.0 & TF & Ir & \\
NGC\,2894 & 092930.2+074304 & 1966 & 2469 & 13.23 & 395~~~ & 42.3 & tf & Sa & NGC\,2894 \\
SDSS\,J09294 & 092946.2+080236 & 1934 & 2435 & 18.59 & ~~~ & & & Ir & NGC\,2894 \\
AGC\,192137 & 092951.8+115536 & 1461 & 1937 & 17.34 & 120~~~ & 39.2 & TFb & Sm & \\
IC\,0540 & 093010.3+075409 & 1856 & 2358 & 14.72 & 256~~~ & 40.4 & TF & Sab & NGC\,2894 \\
KDG~56 & 093012.8+195923 & 441 & 859 & 17.0 & 25~~~ & 8.9 & mem & Ir & NGC\,2903 \\
PGC\,1324298 & 093155.9+073210 & 2036 & 2542 & 18.25 & ~~~ & & & dE & NGC\,2894 \\
NGC\, 2906 & 093206.2+082630 & 1963 & 2463 & 13.40 & 312~~~ & 38.4 & tf & Sbc & NGC\,2894 \\
AGC\,192607 & 093222.2+071808 & 1938 & 2445 & 18.4 & 33~~~ & & & Ir & NGC\,2894 \\
PGC\,027228 & 093444.7+062532 & 366 & 880 & 15.24 & 89~~~ & 13.0 & tf & Im & \\
LEO-T & 093453.5+170304 & -97 & 346 & 16.5 & ~~~ & 0.42 & rgb & Ir & Milky Way \\
UGC\,05107 & 093507.4+050712 & 1806 & 2328 & 15.53 & 174~~~ & 37.4 & tf & Sd & NGC\,2962 \\
AGC\,198335 & 093704.4+095759 & 1347 & 1841 & 19.0 & 53~~~ & & & Ir & \\
AGC\,198430 & 093723.5+040555 & 1820 & 2349 & 18.4 & 46~~~ & 27.2 & TF & Ir & NGC\,2962 \\
SDSS\,J09385 & 093857.1+004134 & 1885 & 2433 & 16.8 & 66~~~ & 37.0 & TF & Sm & \\
AGC\,192830 & 093922.3+045708 & 1703 & 2229 & 16.4 & 167~~~ & & & BCD & NGC\,2962 \\
AGC\,192937 & 094021.1+044406 & 1788 & 2316 & 18.08 & 44~~~ & 30.5 & TF & BCD & NGC\,2962 \\
IC\,0549 & 094043.2+035733 & 1125 & 1657 & 15.17 & 117~~~ & 22.7 & TF & Im & \\
NGC\,2962 & 094053.9+050957 & 1775 & 2301 & 12.91 & 414~~~ & 33.4 & SN & S0a & NGC\,2962 \\
AGC\,192833 & 094056.3+050241 & 1679 & 2205 & 17.2 & 49~~~ & & & Ir & NGC\,2962 \\
PGC\,1175027 & 094117.0+004616 & 1755 & 2304 & 18.07 & ~~~ & & & BCD & NGC\,2967 \\
NGC\,2967 & 094203.3+002011 & 1682 & 2234 & 12.28 & 131~~~ & & & Sc & NGC\,2967 \\
NGC\,2966 & 094211.5+044024 & 1850 & 2379 & 14.11 & 243~~~ & 25.8 & tf & Sbc & NGC\,2962 \\
SDSS\,J09421 & 094218.9+044122 & 1842 & 2371 & 18.52 & ~~~ & & & dE & NGC\,2962 \\
AGC\,193813 & 094250.9+045324 & 1746 & 2274 & 17.2 & 87~~~ & 38.4 & TF & Ir & NGC\,2962 \\
AGC\,198337 & 094251.2+093800 & 1290 & 1790 & 19.0 & 34~~~ & 22.5 & TF & Ir & \\
AGC\,192835 & 094302.2+050144 & 1771 & 2299 & 18.4 & 95~~~ & & & Sm & NGC\,2962 \\
AGC\,193802 & 094419.9+100331 & 1303 & 1801 & 18.6 & 43~~~ & 27.5 & TF & Ir & SDSS\,0944 \\
SDSS\,J09443 & 094437.1+100046 & 1313 & 1811 & 17.1 & 62~~~ & 23.6 & TF & Ir & SDSS\,0944 \\
IC\,0559 & 094443.8+093655 & 370 & 871 & 14.98 & 67~~~ & 9.4 & tf & BCD & \\
AGC\,191869 & 094458.9+082212 & 1577 & 2086 & 16.49 & ~~~ & & & BCD & \\
SDSS\,J09450 & 094503.8+011350 & 1706 & 2255 & 18.62 & ~~~ & & & BCD & UGC\,5228 \\
SDSS\,J09454 & 094541.0+013704 & 1733 & 2281 & 18.63 & ~~~ & & & BCD & UGC\,5228 \\
UGC\,05224 & 094552.1+025839 & 1693 & 2234 & 15.81 & 140~~~ & 31.4 & TF & Sd & \\
IC\,560 & 094553.4-001606 & 1639 & 2196 & 14.01 & ~~~ & & & S0 & NGC\,2967 \\
UGC\,05228 & 094603.6+014006 & 1662 & 2210 & 13.97 & 268~~~ & 31.9 & tf & Scd & UGC\,5228 \\
PGC\,1143397 & 094628.6-002603 & 1599 & 2157 & 16.3 & ~~~ & & & dEem& NGC\,2967 \\
AGC\,198456 & 094642.4+070807 & 1703 & 2220 & 18.9 & 57~~~ & & & Ir & \\
UGC\,05238 & 094654.1+003029 & 1566 & 2120 & 15.43 & 214~~~ & 35.8 & tf & Sd & NGC\,2967 \\
UGC\,05242 & 094705.5+005752 & 1630 & 2182 & 16.95 & 114~~~ & 33.4 & TF & Sm & UGC\,5228 \\
UGC\,05249 & 094745.4+023738 & 1672 & 2216 & 14.55 & ~~~ & 27.7 & tf & Sdm & \\
PGC\,1219995 & 094759.5+021322 & 1737 & 2283 & 17.58 & ~~~ & & & BCD & \\
AGC\,191803 & 094805.9+070744 & 352 & 870 & 16.5 & 55~~~ & 14.9 & tf & Im & \\
PGC\,1145436 & 094842.3-002115 & 1684 & 2243 & 17.4 & ~~~ & & & dEem& NGC\,2967 \\
AGC\,193921 & 094914.9+154827 & 1307 & 1768 & 19.0 & 39~~~ & & & Ir & \\
NGC\,3018 & 094941.4+003715 & 1652 & 2207 & 13.89 & ~~~ & & & Scd & NGC\,3023 \\
NGC\,3023 & 094952.6+003705 & 1668 & 2223 & 13.51 & 133~~~ & 31.5 & tf & Sc & NGC\,3023 \\
NGC\,3020 & 095006.6+124848 & 1284 & 1767 & 13.45 & 215~~~ & 29.4 & tf & Scd & NGC\,3020 \\
NGC\,3024 & 095027.4+124556 & 1259 & 1742 & 14.07 & 245~~~ & 31.1 & TF & Sc & NGC\,3020 \\
SDSS\,J09503 & 095031.3+002427 & 1694 & 2250 & 19.86 & ~~~ & & & Ir & NGC\,3023 \\
AGC\,192239 & 095036.2+124833 & 1178 & 1661 & 17.24 & ~~~ & & & BCD & NGC\,3020 \\
UGC\,05288 & 095117.0+074942 & 378 & 894 & 14.44 & 93~~~ & 11.41 & rgb & Sm & \\
LSBC\,L1-47 & 095138.6+002210 & 1685 & 2242 & 17.40 & 62~~~ & 26.1 & TF & Ir & NGC\,3023 \\
DDO~65 & 095144.1+012655 & 1637 & 2189 & 16.3 & 94~~~ & 29.3 & TF & Ir & NGC\,3023 \\
NGC\,3041 & 095307.1+164040 & 1271 & 1727 & 12.30 & ~~~ & 26.4 & tf & Sc & \\
NGC\,3044 & 095340.9+013447 & 1082 & 1634 & 12.47 & 332~~~ & 22.8 & tf & Scd & NGC\,3044 \\
PGC\,135729 & 095359.1+020017 & 1551 & 2101 & 18.10 & ~~~ & & & Ir & \\
SDSS\,J09535 & 095359.4+025209 & 1574 & 2120 & 17.96 & ~~~ & & & Ir & \\
PGC\,135730 & 095404.5+013223 & 1168 & 1721 & 19.0 & ~~~ & & & Ir & NGC\,3044 \\
SDSS\,J09540 & 095407.3+092135 & 1291 & 1799 & 17.11 & ~~~ & & & BCD & NGC\,3049 \\
AGC\,192423 & 095430.5+095212 & 1318 & 1822 & 17.95 & 40~~~ & & & BCD & NGC\,3049 \\
HIPASSJ095 & 095427.8+015548 & 1603 & 2154 & 16.5 & 127~~~ & 27.8 & TFb & Sdm & \\
AGC\,192959 & 095435.7+042308 & 1579 & 2117 & 17.67 & 77~~~ & 38.9 & TF & Ir & NGC\,3055 \\
PGC\,1200167 & 095445.0+013634 & 1751 & 2303 & 17.53 & ~~~ & & & Sm & \\
NGC\,3049 & 095449.6+091617 & 1283 & 1791 & 13.67 & 203~~~ & 30.2 & TF & Sb & NGC\,3049 \\
NGC\,3055 & 095518.0+041612 & 1610 & 2149 & 12.65 & 266~~~ & 31.6 & tf & Sc & NGC\,3055 \\
UGCA\,188 & 095529.6+082327 & 1097 & 1611 & 15.59 & 101~~~ & 25.9 & TF & Sm & \\
AGC\,192960 & 095537.8+042836 & 1778 & 2316 & 17.17 & 61~~~ & 24.1 & TF & Sm & \\
UGC\,05332 & 095548.2+162449 & 659 & 1119 & 16.47 & 52~~~ & 16.6 & TF & Im & \\
UGC\,05347 & 095716.5+043136 & 1962 & 2500 & 15.30 & 211~~~ & 39.4 & TF & Scd & \\
AGC\,198437 & 095724.2+053942 & 1979 & 2511 & 19.0 & 72~~~ & & & Ir & \\
LSBC\,L1-099 & 095828.8+014141 & 1610 & 2164 & 17.14 & 93~~~ & 21.2 & TFb & Sm & Ark~227 \\
PGC\,1155688 & 095830.2+000243 & 1752 & 2314 & 17.91 & ~~~ & & & BCD & \\
LSBC\,L1-100 & 095846.8+022050 & 1526 & 2076 & 17.90 & 98~~~ & & & Sm & \\
SEX~B & 100000.0+051956 & 110 & 645 & 11.92 & 37~~~ & 1.36 & rgb & Ir & \\
PGC\,1209966 & 100005.8+015440 & 1679 & 2232 & 17.86 & ~~~ & & & Ir & \\
Ark~227 & 100010.4+020922 & 1589 & 2141 & 14.82 & ~~~ & & & dE & Ark~227 \\
UGC\,05376 & 100027.1+032228 & 1855 & 2401 & 14.29 & 338~~~ & 45.3 & TFb & Scd & UGC\,5376 \\
UGC\,05377 & 100031.6+031219 & 1946 & 2493 & 15.24 & ~~~ & & & Sm & UGC\,5376 \\
SDSS\,J10005 & 100059.1+032752 & 1763 & 2308 & 17.02 & ~~~ & & & Sm & UGC\,5376 \\
AGC\,202171 & 100109.5+084656 & 993 & 1507 & 18.02 & ~~~ & & & BCD & \\
RFGC\,1688 & 100110.3+005432 & 1135 & 1694 & 17.84 & ~~~ & & & Sdm & \\
PGC\,3121233 & 100153.8+022450 & 1823 & 2374 & 18.20 & ~~~ & & & Im & UGC\,5376 \\
AGC\,204045 & 100200.0+044728 & 1512 & 2050 & 17.82 & ~~~ & & & BCD & \\
PGC\,3279243 & 100227.1+021001 & 1651 & 2204 & 18.71 & ~~~ & & & Ir & \\
UGC\,05401 & 100231.3+190159 & 1906 & 2348 & 16.30 & 120~~~ & 40.0 & TF & Sm & UGC\,5403 \\
UGC\,05403 & 100235.5+191037 & 1958 & 2399 & 14.45 & 266~~~ & 46.4 & TF & S0em& UGC\,5403 \\
PGC\,3279188 & 100315.4+020543 & 1544 & 2098 & 17.65 & ~~~ & & & Im & \\
Mrk~714 & 100408.7+063038 & 944 & 1473 & 15.81 & 42~~~ & 10.5 & TF & BCD & \\
PGC\,1230703 & 100425.1+023331 & 914 & 1466 & 18.47 & ~~~ & & & Ir & \\
PGC\,1201224 & 100517.6+013828 & 1060 & 1617 & 17.26 & ~~~ & & & Ir & NGC\,3166 \\
AGC\,202297 & 100603.8+103816 & 1394 & 1898 & 17.0 & 258~~~ & & & Sc & \\
AGC\,205108 & 100640.3+121900 & 1329 & 1822 & 19.24 & 26~~~ & & & Ir & \\
SEGUE~1 & 100703.6+160440 & 67 & 533 & 16.2 & ~~~ & 0.02 & rgb & Sph & Milky Way \\
AGC\,203862 & 100704.5+050025 & 1534 & 2073 & 17.3 & 34~~~ & 11.3 & TF & Ir & \\
UGC\,05453 & 100707.2+155902 & 700 & 1167 & 15.39 & 53~~~ & & & Im & \\
UGC\,05456 & 100719.8+102143 & 360 & 866 & 13.74 & 61~~~ & 5.60 & rgb & Im & \\
SDSS\,J10072 & 100724.1+051931 & 1414 & 1951 & 18.05 & ~~~ & & & Ir & \\
PGC\,1223942 & 100806.9+022043 & 1760 & 2314 & 16.52 & ~~~ & & & BCD & \\
PGC\,029471 & 100810.3+022748 & 1873 & 2426 & 15.55 & ~~~ & & & dE & \\
Leo~I & 100828.0+121823 & 125 & 619 & 11.16 & ~~~ & 0.25 & rgb & Sph & Milky Way \\
FGC\,120A & 100917.4+052414 & 1512 & 2050 & 17.5 & 83~~~ & 37.5 & TFb & Sd & \\
SDSS\,J10102 & 101020.5+074513 & 1090 & 1614 & 18.12 & ~~~ & & & Ir & \\
NGC\,3156 & 101241.3+030746 & 1140 & 1691 & 13.07 & 200~~~ & 22.5 & sbf & S0em& NGC\,3166 \\
UGC\,05504 & 101249.1+070612 & 1364 & 1893 & 16.25 & 147~~~ & 30.3 & tf & Sd & \\
NGC\,3165 & 101331.3+032230 & 1145 & 1695 & 14.50 & 128~~~ & 17.9 & TF & Sm & NGC\,3166 \\
PGC\,3282143 & 101332.3+010601 & 1952 & 2514 & 19.0 & ~~~ & & & Ir & \\
NGC\,3166 & 101345.7+032530 & 1148 & 1698 & 11.42 & 193~~~ & & & S0a & NGC\,3166 \\
UGC\,05522 & 101358.9+070126 & 1037 & 1566 & 14.61 & 211~~~ & 31.6 & TFb & Scd & \\
NGC\,3169 & 101414.9+032759 & 1041 & 1591 & 11.25 & 452~~~ & 18.8 & SN & Sa & NGC\,3166 \\
UGC\,05539 & 101555.0+024109 & 1081 & 1636 & 16.10 & 141~~~ & 27.0 & TFb & Sm & NGC\,3166 \\
KKH~60 & 101559.5+064816 & 1438 & 1969 & 17.84 & 94~~~ & 30.6 & TFb & Im & \\
SDSS\,J10165 & 101659.0+034235 & 1033 & 1582 & 17.37 & ~~~ & & & Im & NGC\,3166 \\
PGC\,1256137 & 101702.3+033846 & 862 & 1412 & 17.12 & 33~~~ & & & BCD & \\
PGC\,213680 & 101709.0+042040 & 1115 & 1661 & 17.36 & ~~~ & & & Ir & NGC\,3166 \\
UGC\,05551 & 101711.9+041949 & 1148 & 1694 & 16.8 & 56~~~ & 17.8 & TF & Im & NGC\,3166 \\
AGC\,208392 & 101803.7+041835 & 1130 & 1676 & 19.0 & 34~~~ & & & Ir & NGC\,3166 \\
PGC\,030133 & 101901.5+211702 & 972 & 1401 & 15.4 & 78~~~ & 24.8 & TF & Sm & NGC\,3227 \\
SDSS\,J10190 & 101904.6+171100 & 1846 & 2307 & 18.12 & ~~~ & & & BCD & \\
NGC\,3213 & 102117.3+193906 & 1227 & 1670 & 14.17 & ~~~ & 32.0 & tf & Sd & \\
PGC\,1178576 & 102138.9+005400 & 495 & 1060 & 17.5 & ~~~ & & & Ir & \\
LEO-P & 102144.8+180520 & 135 & 590 & 17.2 & ~~~ & 2.0 & txt & Im & \\
PGC\,1609953 & 102322.5+195452 & 1081 & 1522 & 17.32 & ~~~ & & & dE & NGC\,3227 \\
NGC\,3226 & 102327.0+195354 & 1197 & 1638 & 12.34 & ~~~ & 23.7 & sbf & E & NGC\,3227 \\
NGC\,3227 & 102330.6+195154 & 1039 & 1480 & 11.55 & 400~~~ & 22.2 & TF & Sab & NGC\,3227 \\
UGC\,05633 & 102440.1+144523 & 1240 & 1720 & 14.46 & 167~~~ & 36.2 & TF & Sm & UGC\,5646 \\
NGC\,3239 & 102504.8+170949 & 623 & 1086 & 11.70 & ~~~ & 7.9 & tf & Im & \\
AGC\,203913 & 102546.4+053909 & 966 & 1506 & 16.17 & 99~~~ & 28.4 & TF & Sd & \\
UGC\,05646 & 102553.1+142148 & 1225 & 1708 & 14.20 & 221~~~ & 27.6 & TF & Sbc & UGC\,5646 \\
IC\,610 & 102628.4+202859 & 1054 & 1491 & 14.72 & 268~~~ & 31.8 & tf & Sbc & NGC\,3227 \\
NGC\,3246 & 102641.8+035143 & 1961 & 2511 & 13.91 & 244~~~ & 35.5 & tf & Sd & NGC\,3246 \\
AGC\,208295 & 102827.2+081026 & 1317 & 1842 & 18.9 & 91~~~ & & & Ir & \\
UGC\,05675 & 102830.0+193345 & 983 & 1427 & 16.66 & 72~~~ & 27.3 & TF & Sm & NGC\,3227 \\
UGC\,05677 & 102838.2+033338 & 963 & 1515 & 15.07 & ~~~ & 22.0 & tf & Sd & \\
PGC\,1214845 & 102843.0+020349 & 1860 & 2420 & 17.66 & ~~~ & & & BCD & \\
AGC\,208394 & 102843.8+044404 & 992 & 1538 & 19.0 & 27~~~ & & & Ir & \\
VIII Zw081 & 102848.1+041403 & 1967 & 2515 & 15.76 & 104~~~ & 26.9 & TF & BCD & NGC\,3246 \\
AGC\,202218 & 102855.8+095144 & 1024 & 1538 & 16.71 & 39~~~ & 11.7 & TF & Im & \\
SDSS\,J10304 & 103044.3+060734 & 458 & 996 & 16.80 & 27~~~ & 7.8 & tf & Ir & \\
AGC\,205156 & 103052.9+122648 & 762 & 1259 & 18.6 & 21~~~ & 10.4 & tf & BCD & NGC\,3379 \\
UGC\,05708 & 103113.2+042819 & 987 & 1534 & 14.37 & 169~~~ & 21.3 & TF & Sd & UGC\,5708 \\
SDSS\,J10313 & 103137.3+043422 & 1013 & 1560 & 16.27 & ~~~ & & & BCD & UGC\,5708 \\
AGC\,202244 & 103140.8+135005 & 1141 & 1629 & 16.5 & 102~~~ & 34.3 & TF & Im & \\
AGC\,204139 & 103201.3+042046 & 957 & 1505 & 18.42 & 68~~~ & & & Ir & \\
SDSS\,J10331 & 103316.2+181311 & 1107 & 1562 & 17.57 & ~~~ & & & Ir & \\
AGC\,202016 & 103319.2+101122 & 1270 & 1783 & 19.10 & 32~~~ & & & Ir & \\
AGC\,205161 & 103405.5+154650 & 1081 & 1555 & 17.90 & 114~~~ & & & Ir & \\
NGC\,3279 & 103442.8+111149 & 1236 & 1742 & 13.93 & 347~~~ & 32.2 & tf & Scd & \\
AGC\,202248 & 103456.1+112932 & 1020 & 1524 & 17.5 & 62~~~ & 10.4 & mem & Ir & NGC\,3379 \\
LeG\,03 & 103548.9+082853 & 987 & 1511 & 17.3 & 70~~~ & 10.4 & mem & Sm & NGC\,3379 \\
NGC\,3299 & 103624.0+124224 & 453 & 949 & 13.30 & 112~~~ & 10.4 & mem & Sm & NGC\,3379 \\
AGC\,205165 & 103704.8+152015 & 586 & 1063 & 16.4 & 27~~~ & 10.4 & mem & Im & NGC\,3379 \\
FGC\,125a & 103728.7+122346 & 1178 & 1676 & 17.4 & 59~~~ & 25.0 & TF & Sd & \\
AGC\,200499 & 103808.0+102251 & 1009 & 1521 & 14.40 & 178~~~ & 36.5 & TF & BCD & \\
AGC\,208397 & 103858.1+035227 & 573 & 1124 & 19.7 & 33~~~ & 26.2 & tf & Ir & \\
UGC\,05797 & 103925.2+014307 & 511 & 1074 & 14.30 & 47~~~ & & & BCD & \\
LeG\,05 & 103942.9+123805 & 629 & 1125 & 16.80 & ~~~ & 10.4 & mem & dE & NGC\,3379 \\
LeG\,06 & 103955.6+135434 & 863 & 1350 & 18.30 & 21~~~ & 10.4 & mem & Ir & NGC\,3379 \\
AGC\,208399 & 104010.7+045432 & 561 & 1106 & 20.0 & 23~~~ & 20.9 & tf & Ir & \\
UGC\,05812 & 104056.5+122818 & 857 & 1355 & 15.50 & 56~~~ & 10.4 & mem & Sm & NGC\,3379 \\
AGC\,205078 & 104126.1+070216 & 999 & 1532 & 19.0 & 32~~~ & & & Ir & \\
AGC\,203080 & 104141.0+134930 & 1127 & 1615 & 17.46 & ~~~ & & & dE & NGC\,3338 \\
FS\,04 & 104200.3+122006 & 621 & 1119 & 15.7 & 36~~~ & 10.4 & mem & Sm & NGC\,3379 \\
NGC\,3338 & 104207.6+134449 & 1157 & 1646 & 11.44 & 339~~~ & 24.9 & tf & Sc & NGC\,3338 \\
AGC\,203082 & 104226.5+135726 & 1133 & 1620 & 17.8 & 41~~~ & 17.5 & TF & Ir & NGC\,3338 \\
UGC\,05832 & 104248.5+132736 & 1070 & 1561 & 14.31 & 102~~~ & 18.6 & TF & Scd & NGC\,3338 \\
AGC\,205268 & 104252.4+134428 & 1001 & 1490 & 17.4 & ~~~ & & & BCD & NGC\,3338 \\
AGC\,200543 & 104305.5+133040 & 1111 & 1601 & 16.2 & 70~~~ & 18.1 & TF & Im & NGC\,3338 \\
NGC\,3346 & 104338.9+145218 & 1135 & 1615 & 12.59 & 162~~~ & & & Scd & NGC\,3338 \\
NGC\,3351 & 104357.7+114213 & 624 & 1127 & 10.60 & 270~~~ & 10.05 & cep & Sb & NGC\,3379 \\
AGC\,205445 & 104435.3+135623 & 490 & 977 & 16.4 & ~~~ & 10.4 & mem & Sph & NGC\,3379 \\
PGC\,1174229 & 104456.3+004427 & 1412 & 1979 & 17.12 & ~~~ & & & Im & \\
LeG\,13 & 104457.5+115458 & 734 & 1235 & 17.36 & 24~~~ & 11.3 & TF & Ir & NGC\,3379 \\
AGC\,205270 & 104509.8+152700 & 1083 & 1559 & 17.14 & 51~~~ & & & BCD & \\
PGC\,3090074 & 104602.8+193216 & 1197 & 1642 & 16.57 & ~~~ & & & dE & \\
NGC\,3365 & 104612.6+014848 & 789 & 1351 & 13.18 & 234~~~ & 18.2 & tf & Sd & \\
LeG\,14 & 104614.2+125737 & 749 & 1243 & 18.70 & ~~~ & 10.4 & mem & Sph & NGC\,3379 \\
LeG\,17 & 104641.3+121938 & 880 & 1378 & 17.0 & ~~~ & 10.4 & mem & Sph & NGC\,3379 \\
NGC\,3368 & 104645.7+114911 & 740 & 1242 & 10.10 & 343~~~ & 10.42 & cep & Sab & NGC\,3379 \\
LeG\,18 & 104653.2+124440 & 488 & 983 & 18.90 & 38~~~ & 10.4 & mem & Ir & NGC\,3379 \\
KK~94 & 104656.8+125957 & 684 & 1178 & 17.5 & ~~~ & 10.4 & mem & Tr & NGC\,3379 \\
LeG\,21 & 104700.8+125735 & 696 & 1190 & 18.60 & ~~~ & 10.4 & mem & Ir & NGC\,3379 \\
NGC\,3370 & 104704.0+171625 & 1153 & 1615 & 12.29 & ~~~ & 27.50 & cep & Sbc & \\
DDO~88 & 104722.4+140412 & 431 & 917 & 14.40 & 46~~~ & 7.73 & rgb & Im & NGC\,3379 \\
NGC\,3377 & 104742.4+135908 & 536 & 1023 & 11.20 & ~~~ & 10.91 & sbf & E & NGC\,3379 \\
NGC\,3379 & 104749.6+123454 & 774 & 1270 & 10.23 & ~~~ & 11.12 & sbf & E & NGC\,3379 \\
NGC\,3384 & 104816.9+123746 & 556 & 1052 & 10.89 & ~~~ & 11.38 & sbf & S0 & NGC\,3379 \\
NGC\,3389 & 104827.9+123159 & 1159 & 1656 & 12.51 & 266~~~ & 32.8 & SN & Sc & \\
AGC\,200596 & 104853.7+140728 & 400 & 885 & 15.66 & ~~~ & & & dE & NGC\,3379 \\
Mrk~1263 & 104856.8+121142 & 1171 & 1670 & 15.70 & 125~~~ & 34.2 & TF & BCD & \\
AGC\,200600 & 104859.7+105007 & 1782 & 2290 & 16.25 & 120~~~ & 38.4 & TF & Sm & \\
UGC\,05923 & 104907.6+065502 & 538 & 1071 & 14.41 & 142~~~ & 22.3 & tf & S0em& \\
AGC\,200603 & 104917.1+122519 & 1234 & 1731 & 15.72 & 68~~~ & 14.2 & TF & Sm & \\
PGC\,032376 & 104918.4+122242 & 1200 & 1698 & 18.0 & ~~~ & & & Ir & NGC\,3379 \\
AGC\,202253 & 104926.7+121528 & 1188 & 1686 & 17.5 & ~~~ & & & BCD & NGC\,3379 \\
AGC\,205197 & 104942.8+134941 & 1190 & 1677 & 19.0 & 42~~~ & & & Ir & \\
AGC\,205313 & 104947.9+123626 & 626 & 1122 & 18.0 & 30~~~ & 12.1 & TF & Im & NGC\,3379 \\
AGC\,205198 & 105001.8+134705 & 1168 & 1656 & 17.18 & 53~~~ & 19.1 & TF & BCD & \\
LSBGL1-134 & 105008.9+011554 & 1405 & 1969 & 17.38 & ~~~ & & & Sm & \\
UGC\,05944 & 105019.0+131621 & 928 & 1419 & 15.27 & ~~~ & 11.07 & sbf & Sph & NGC\,3379 \\
UGC\,05945 & 105025.5+173351 & 1008 & 1468 & 14.7 & 124~~~ & 22.8 & TF & Im & NGC\,3607 \\
UGC\,05947 & 105030.4+193839 & 1137 & 1581 & 14.92 & 70~~~ & 10.4 & TF & Im & \\
UGC\,05948 & 105038.2+154548 & 987 & 1460 & 16.65 & 106~~~ & 26.3 & TFb & Im & \\
NGC\,3412 & 105053.3+132444 & 697 & 1187 & 11.44 & ~~~ & 11.3 & sbf & S0 & NGC\,3379 \\
NGC\,3423 & 105114.4+055024 & 825 & 1364 & 11.61 & 156~~~ & 19.5 & TF & Scd & NGC\,3423 \\
LeG\,26 & 105121.1+125057 & 483 & 977 & 16.9 & ~~~ & 10.4 & mem & Sph & NGC\,3379 \\
AGC\,205540 & 105131.3+140653 & 691 & 1176 & 18.0 & ~~~ & 10.4 & mem & Ir & NGC\,3379 \\
KKH~64 & 105132.1+032718 & 881 & 1433 & 16.5 & 74~~~ & 23.1 & TF & Im & NGC\,3423 \\
UGC\,05974 & 105135.0+043459 & 859 & 1405 & 14.82 & ~~~ & 27.2 & tf & Sd & NGC\,3423 \\
SDSS\,J10514 & 105148.7+194606 & 1276 & 1719 & 18.24 & ~~~ & & & BCD & \\
AGC\,205544 & 105204.7+150149 & 692 & 1171 & 17.1 & ~~~ & 10.4 & mem & Sph & NGC\,3379 \\
AGC\,202456 & 105219.5+110235 & 669 & 1175 & 16.2 & ~~~ & 10.4 & mem & Sph & NGC\,3379 \\
PGC\,032630 & 105221.4+175607 & 1182 & 1639 & 15.20 & ~~~ & & & dS0 & \\
UGC\,05989 & 105231.8+194732 & 1016 & 1458 & 14.41 & 126~~~ & 16.1 & TF & Sm & \\
SDSS\,J10523 & 105234.9+170842 & 928 & 1391 & 18.26 & ~~~ & & & Ir & NGC\,3607 \\
MGC\,0013223 & 105240.6-000116 & 1569 & 2139 & 17.8 & ~~~ & & & Ir & PGC\,032664 \\
PGC\,032664 & 105248.6+000204 & 1607 & 2177 & 15.75 & 89~~~ & 20.8 & TF & BCD & PGC\,032664 \\
NGC\,3443 & 105300.1+173426 & 1009 & 1468 & 14.83 & ~~~ & 22.0 & tf & Scd & NGC\,3607 \\
PGC\,135768 & 105303.3+022937 & 854 & 1411 & 17.46 & ~~~ & & & Im & PGC\,032687 \\
SDSS\,J10531 & 105314.5+175028 & 1052 & 1509 & 17.57 & ~~~ & & & Im & NGC\,3607 \\
PGC\,032687 & 105318.9+023736 & 866 & 1423 & 15.95 & 78~~~ & 18.6 & TF & Sm & PGC\,032687 \\
NGC\,3447 & 105323.9+164630 & 940 & 1405 & 14.46 & ~~~ & & & Sm & NGC\,3607 \\
NGC\,3447b & 105329.6+164710 & 971 & 1436 & 14.3 & ~~~ & & & Im & NGC\,3607 \\
UGC\,06014 & 105342.7+094339 & 965 & 1480 & 15.2 & 94~~~ & 17.0 & TF & Sm & \\
SDSS\,J10540 & 105400.0+094952 & 1023 & 1537 & 17.54 & ~~~ & & & BCD & \\
UGC\,06022 & 105415.4+174837 & 848 & 1305 & 16.4 & 86~~~ & 31.8 & TF & Sm & NGC\,3607 \\
NGC\,3454 & 105429.2+172040 & 977 & 1438 & 13.71 & ~~~ & 27.6 & tf & Sc & NGC\,3607 \\
NGC\,3455 & 105431.1+171705 & 978 & 1439 & 14.31 & ~~~ & 29.0 & tf & Sb & NGC\,3607 \\
NGC\,3457 & 105448.6+173716 & 1025 & 1484 & 13.6 & ~~~ & 20.7 & sbf & E & NGC\,3607 \\
AGC\,202033 & 105503.6+140515 & 1968 & 2453 & 18.9 & ~~~ & & & Ir & \\
PGC\,1533359 & 105506.7+172746 & 1035 & 1495 & 16.90 & ~~~ & & & Im & NGC\,3607 \\
UGC\,06035 & 105529.0+170830 & 947 & 1409 & 15.44 & ~~~ & & & Im & NGC\,3607 \\
PGC\,032833 & 105539.2+022345 & 829 & 1386 & 16.48 & 72~~~ & 21.7 & TF & Im & \\
PGC\,032843 & 105544.4+170018 & 1018 & 1481 & 15.32 & ~~~ & & & BCD & NGC\,3607 \\
LSBC\,D640-1 & 105555.8+122019 & 699 & 1196 & 18.40 & 22~~~ & 10.4 & mem & Ir & NGC\,3379 \\
Mrk~1271 & 105609.1+061022 & 837 & 1373 & 14.81 & 128~~~ & 23.3 & TF & BCD & NGC\,3423 \\
AGC\,202035 & 105613.9+120040 & 840 & 1339 & 16.9 & 30~~~ & 10.4 & mem & Im & NGC\,3379 \\
SDSS\,J10561 & 105619.9+170506 & 830 & 1293 & 18.52 & ~~~ & & & Im & NGC\,3607 \\
SDSS\,J10563 & 105638.6+172301 & 818 & 1278 & 18.20 & ~~~ & & & BCD & NGC\,3607 \\
AGC\,202260 & 105738.2+135844 & 1078 & 1563 & 17.50 & 92~~~ & 28.8 & TFb & Im & \\
CGCG\,095-78 & 105802.2+193019 & 538 & 982 & 15.6 & 62~~~ & 11.7 & TF & Im & \\
AGC\,205278 & 105852.4+140747 & 548 & 1032 & 17.3 & 36~~~ & 11.8 & TF & Ir & NGC\,3379 \\
NGC\,3485 & 110002.4+145029 & 1301 & 1779 & 12.67 & 135~~~ & & & Sb & \\
NGC\,3489 & 110018.6+135404 & 538 & 1023 & 11.06 & 113~~~ & 12.08 & sbf & S0a & NGC\,3379 \\
UGC\,06083 & 110023.8+164132 & 811 & 1276 & 15.29 & 146~~~ & 27.2 & TF & Sbc & NGC\,3607 \\
SDSS\,J11004 & 110047.2+165256 & 1022 & 1485 & 18.13 & ~~~ & & & Im & NGC\,3607 \\
UGC\,06095 & 110104.4+190600 & 1002 & 1448 & 16.31 & 94~~~ & 29.4 & TF & Im & NGC\,3607 \\
NGC\,3495 & 110116.2+033741 & 944 & 1494 & 12.42 & ~~~ & 18.5 & tf & Scd & \\
UGC\,06112 & 110235.2+164406 & 909 & 1373 & 14.79 & 152~~~ & 25.7 & TF & Sd & NGC\,3607 \\
NGC\,3501 & 110247.3+175922 & 1012 & 1466 & 13.61 & ~~~ & 24.3 & tf & Scd & NGC\,3607 \\
AGC\,202040 & 110301.8+080253 & 1194 & 1717 & 18.10 & 96~~~ & 37.6 & TFb & Ir & \\
NGC\,3507 & 110325.4+180810 & 862 & 1315 & 12.07 & ~~~ & & & Sb & NGC\,3607 \\
AGC\,215256 & 110326.4+160059 & 1102 & 1571 & 16.88 & 105~~~ & 37.5 & TF & Ir & \\
AGC\,219117 & 110346.7+083419 & 1575 & 2095 & 18.72 & 68~~~ & & & Ir & \\
AGC\,210023 & 110426.4+114522 & 629 & 1128 & 16.4 & 44~~~ & 10.3 & TF & Im & NGC\,3379 \\
KK~SG~20 & 110440.2+000329 & 636 & 1203 & 17.5 & 25~~~ & 10.7 & mem & Ir & NGC\,3521 \\
SDSS\,J11045 & 110456.8+173830 & 798 & 1254 & 17.43 & ~~~ & & & dEem& NGC\,3607 \\
PGC\,033523 & 110532.5+173823 & 898 & 1354 & 14.89 & ~~~ & & & dEem& NGC\,3607 \\
UGC\,6145 & 110535.0-015149 & 546 & 1122 & 16.50 & 41~~~ & 10.7 & TF & Ir & NGC\,3521 \\
NGC\,3521 & 110548.6-000209 & 598 & 1165 & 9.80 & 441~~~ & 10.7 & TF & Sbc & NGC\,3521 \\
UGC\,06151 & 110556.3+194931 & 1227 & 1666 & 14.90 & ~~~ & & & Sm & \\
AGC\,213757 & 110559.6+072225 & 1472 & 1999 & 17.49 & 57~~~ & & & BCD & \\
PGC\,1558217 & 110627.3+182324 & 1130 & 1580 & 17.46 & 66~~~ & 30.4 & TF & Sdm & \\
NGC\,3524 & 110632.1+112307 & 1216 & 1717 & 13.36 & ~~~ & & & S0 & \\
AGC\,215262 & 110635.3+121348 & 1461 & 1956 & 18.2 & 63~~~ & & & Ir & \\
SDSS\,J11065 & 110651.1+173003 & 830 & 1287 & 18.52 & ~~~ & & & Sph & NGC\,3607 \\
NGC\,3526 & 110656.7+071026 & 1259 & 1787 & 13.86 & 196~~~ & 20.8 & TF & Scd & \\
UGC\,06169 & 110703.4+120336 & 1405 & 1901 & 14.56 & 241~~~ & 34.6 & TF & Sbc & \\
UGC\,06171 & 110710.1+183412 & 1092 & 1541 & 15.13 & 146~~~ & 29.5 & TF & Sdm & NGC\,3607 \\
UGC\,06181 & 110746.6+193258 & 1060 & 1501 & 15.54 & ~~~ & & & Sm & NGC\,3607 \\
PGC\,033816 & 110923.2+105003 & 1404 & 1908 & 15.27 & 66~~~ & & & Sm & NGC\,3547 \\
NGC\,3547 & 110955.9+104314 & 1438 & 1942 & 13.20 & 204~~~ & 28.7 & tf & Sbc & NGC\,3547 \\
AGC\,210111 & 111025.2+100733 & 1167 & 1675 & 15.98 & 60~~~ & 18.7 & TF & Ir & \\
AGC\,213064 & 111054.5+093719 & 1414 & 1925 & 15.52 & 124~~~ & 31.4 & TF & BCD & \\
PGC\,033959 & 111054.9+010531 & 802 & 1362 & 15.88 & 71~~~ & 16.2 & TF & Im & \\
UGC\,06233 & 111128.3+065427 & 1398 & 1926 & 14.66 & 212~~~ & 34.8 & TF & S0em& \\
SDSS\,J11114 & 111147.0+185126 & 856 & 1301 & 17.72 & ~~~ & & & Sm & NGC\,3607 \\
KK~98 & 111215.7+164514 & 1092 & 1553 & 17.40 & ~~~ & & & Tr & NGC\,3607 \\
PGC\,087259 & 111231.7+161723 & 1127 & 1591 & 16.89 & ~~~ & & & dE & \\
IC\,0676 & 111239.8+090321 & 1272 & 1786 & 13.52 & 177~~~ & 20.0 & TF & S0em& \\
UGC\,06248 & 111251.8+101200 & 1134 & 1641 & 16.7 & 26~~~ & & & Sm & \\
AGC\,213796 & 111252.7+075519 & 1242 & 1763 & 17.5 & 78~~~ & 25.6 & TF & BCD & PGC\,34135 \\
PGC\,034135 & 111300.2+075143 & 1233 & 1754 & 15.28 & 118~~~ & 22.7 & TF & Sdm & PGC\,34135 \\
AGC\,215280 & 111316.3+152428 & 1351 & 1822 & 18.2 & 93~~~ & & & Ir & \\
PGC\,1257521 & 111350.6+034342 & 2081 & 2626 & 17.7 & ~~~ & & & BCD & \\
AGC\,215240 & 111350.8+095739 & 1457 & 1965 & 18.55 & 34~~~ & 19.5 & TF & BCD & \\
AGC\,219197 & 111355.2+040619 & 1430 & 1973 & 16.80 & 63~~~ & 20.3 & TF & BCD & \\
PGC\,1219195 & 111405.2+021155 & 1179 & 1732 & 17.12 & ~~~ & & & Sm & NGC\,3640 \\
AGC\,215282 & 111425.2+153202 & 731 & 1200 & 16.84 & 27~~~ & & & Im & \\
NGC\,3592 & 111427.3+171536 & 1180 & 1636 & 14.46 & ~~~ & 25.7 & tf & Sc & \\
NGC\,3593 & 111437.0+124904 & 489 & 977 & 11.86 & 254~~~ & 10.8 & tf & S0a & NGC\,3627 \\
AGC\,202256 & 111445.0+123851 & 490 & 980 & 17.50 & 42~~~ & 11.0 & tf & Ir & NGC\,3627 \\
NGC\,3596 & 111506.2+144713 & 1063 & 1537 & 11.79 & 118~~~ & & & Sc & NGC\,3596 \\
AGC\,215281 & 111516.2+144155 & 962 & 1437 & 18.1 & ~~~ & & & Ir & NGC\,3596 \\
NGC\,3599 & 111527.0+180637 & 726 & 1175 & 12.88 & ~~~ & 20.4 & sbf & S0 & NGC\,3607 \\
AGC\,215284 & 111532.4+143438 & 1002 & 1478 & 17.84 & 23~~~ & & & Ir & NGC\,3596 \\
AGC\,212132 & 111626.1+042011 & 932 & 1473 & 15.95 & 155~~~ & 39.8 & TF & Sdm & \\
PGC\,034407 & 111635.3+180706 & 828 & 1277 & 15.40 & ~~~ & & & S0 & NGC\,3607 \\
PGC\,1224534 & 111642.1+022149 & 1415 & 1966 & 18.13 & ~~~ & & & dE & \\
NGC\,3605 & 111646.7+180102 & 548 & 998 & 13.16 & ~~~ & 20.5 & sbf & E & NGC\,3607 \\
UGC\,06296 & 111651.1+174754 & 862 & 1313 & 14.18 & ~~~ & 24.4 & tf & Scd & NGC\,3607 \\
NGC\,3607 & 111654.5+180307 & 829 & 1278 & 10.93 & ~~~ & 22.8 & sbf & S0 & NGC\,3607 \\
NGC\,3608 & 111658.9+180855 & 1113 & 1562 & 11.57 & ~~~ & 22.9 & sbf & E & NGC\,3607 \\
IC\,2684 & 111701.0+130557 & 451 & 937 & 16.2 & 25~~~ & 10.3 & mem & Tr & NGC\,3627 \\
AGC\,215186 & 111701.2+043944 & 1279 & 1818 & 18.51 & 66~~~ & & & Ir & NGC\,3640 \\
AGC\,215241 & 111702.7+100836 & 1614 & 2119 & 17.72 & 120~~~ & 44.4 & TFb & Sdm & \\
UGC\,06300 & 111717.0+161938 & 948 & 1410 & 15.77 & ~~~ & & & dEem& NGC\,3607 \\
UGC\,06306 & 111727.4+043616 & 1579 & 2118 & 17.4 & 108~~~ & & & Ir & NGC\,3611 \\
NGC\,3611 & 111730.1+043319 & 1447 & 1986 & 12.85 & 375~~~ & & & Sa & NGC\,3611 \\
PGC\,034493 & 111738.2+174905 & 1181 & 1632 & 15.56 & ~~~ & & & S0a & \\
PGC\,1513499 & 111748.4+163824 & 851 & 1311 & 17.18 & ~~~ & & & Im & \\
PGC\,034522 & 111758.0+172629 & 691 & 1145 & 15.05 & ~~~ & & & dE & NGC\,3607 \\
AGC\,213006 & 111803.9+101440 & 806 & 1310 & 18.33 & ~~~ & & & Ir & NGC\,3627 \\
IC\,2703 & 111805.1+173858 & 862 & 1314 & 15.70 & ~~~ & & & dE & NGC\,3607 \\
UGC\,06320 & 111817.3+185049 & 1016 & 1459 & 13.83 & ~~~ & & & Sm & NGC\,3607 \\
PGC\,086629 & 111821.4+174151 & 937 & 1389 & 17.6 & 55~~~ & 25.0 & TF & Ir & NGC\,3607 \\
UGC\,06324 & 111822.1+184418 & 958 & 1402 & 14.77 & ~~~ & & & S0e?& NGC\,3607 \\
PGC\,1192339 & 111826.9+012121 & 770 & 1326 & 18.07 & ~~~ & & & Ir & \\
SDSS\,J11185 & 111850.5+034549 & 1513 & 2056 & 18.4 & ~~~ & & & Ir & \\
NGC\,3623 & 111855.8+130535 & 671 & 1156 & 10.14 & 493~~~ & 12.8 & tf & Sa & NGC\,3627 \\
AGC\,215286 & 111912.7+141940 & 867 & 1343 & 18.0 & 28~~~ & 11.7 & TF & Ir & NGC\,3627 \\
AGC\,202257 & 111914.4+115707 & 719 & 1212 & 17.35 & 51~~~ & 11.7 & TFb & Ir & NGC\,3627 \\
AGC\,215354 & 111915.9+141725 & 659 & 1135 & 17.40 & ~~~ & 10.4 & mem & BCD & NGC\,3627 \\
AGC\,213074 & 111928.1+093544 & 843 & 1351 & 17.43 & 51~~~ & 20.6 & TF & Ir & \\
PGC\,034653 & 111933.2+030053 & 1070 & 1617 & 16.0 & ~~~ & & & Sm & NGC\,3640 \\
AGC\,215287 & 111945.1+153008 & 1209 & 1676 & 16.8 & 103~~~ & 40.5 & TF & Sm & \\
PGC\,3288593 & 111954.0+005019 & 1488 & 2046 & 19.09 & ~~~ & & & Ir & \\
UGC\,06341 & 112000.6+181538 & 1530 & 1977 & 16.06 & 88~~~ & 23.4 & TF & Sdm & NGC\,3626 \\
PGC\,1234729 & 112003.1+024123 & 1460 & 2008 & 17.98 & ~~~ & & & Ir & \\
NGC\,3626 & 112003.8+182125 & 1382 & 1828 & 11.81 & ~~~ & 20.00 & sbf & Sa & NGC\,3626 \\
NGC\,3627 & 112015.0+125928 & 590 & 1075 & 9.74 & 369~~~ & 10.28 & cep & Sb & NGC\,3627 \\
UGC\,06345 & 112015.6+023133 & 1419 & 1968 & 14.07 & 107~~~ & 12.3 & TF & Sm & \\
NGC\,3628 & 112016.9+133520 & 709 & 1190 & 9.97 & 458~~~ & 12.2 & tf & Sb & NGC\,3627 \\
NGC\,3630 & 112017.0+025751 & 1317 & 1864 & 12.90 & ~~~ & 28.9 & tf & S0 & NGC\,3640 \\
PGC\,1553459 & 112045.0+181310 & 1364 & 1811 & 16.61 & ~~~ & & & BCD & \\
SDSS\,J11210 & 112106.9+032807 & 1326 & 1870 & 18.0 & ~~~ & & & Ir & NGC\,3640 \\
NGC\,3640 & 112106.9+031405 & 1118 & 1663 & 11.33 & ~~~ & 27.0 & sbf & E & NGC\,3640 \\
NGC\,3641 & 112108.8+031140 & 1600 & 2145 & 14.12 & ~~~ & 26.7 & sbf & E & NGC\,3640 \\
NGC\,3643 & 112125.0+030050 & 1569 & 2115 & 14.65 & ~~~ & & & S0a & NGC\,3640 \\
PGC\,1534499 & 112125.1+173037 & 865 & 1317 & 16.46 & ~~~ & & & dE & NGC\,3607 \\
SDSS\,J11214 & 112140.3+193643 & 1043 & 1478 & 16.55 & ~~~ & & & Ir & NGC\,3607 \\
PGC\,1519262 & 112148.0+165247 & 1399 & 1855 & 17.12 & ~~~ & & & Ir & NGC\,3626 \\
SDSS\,J11215 & 112151.4+032418 & 1044 & 1588 & 17.34 & ~~~ & & & Ir & NGC\,3640 \\
SDSS\,J11220 & 112204.1+033652 & 1259 & 1801 & 18.40 & ~~~ & & & dE & NGC\,3640 \\
SDSS\,J11221 & 112211.1+043942 & 1131 & 1668 & 17.83 & ~~~ & & & dE & NGC\,3640 \\
IC\,2763 & 112218.5+130354 & 1433 & 1917 & 15.15 & 132~~~ & 24.8 & TF & Sdm & \\
AGC\,219200 & 112220.0+035356 & 1124 & 1665 & 19.0 & 28~~~ & & & Ir & NGC\,3640 \\
IC\,2767 & 112223.2+130440 & 942 & 1425 & 17.06 & 92~~~ & & & Im & NGC\,3627 \\
AGC\,213511 & 112223.4+114738 & 1430 & 1922 & 17.6 & 61~~~ & 28.8 & TF & BCD & \\
AGC\,213436 & 112224.0+125846 & 491 & 975 & 16.7 & ~~~ & 10.3 & mem & dEem& NGC\,3627 \\
AGC\,219201 & 112231.4+053129 & 1405 & 1936 & 18.9 & 24~~~ & & & Ir & \\
IC\, 2781 & 112250.7+122041 & 1406 & 1894 & 17.0 & 72~~~ & 27.6 & TF & Im & \\
NGC\,3655 & 112254.7+163524 & 1355 & 1813 & 12.32 & ~~~ & 30.9 & tf & Sc & NGC\,3626 \\
IC\,2782 & 112255.4+132628 & 727 & 1208 & 15.12 & ~~~ & & & Sph & NGC\,3627 \\
AGC\,215290 & 112259.1+122738 & 1475 & 1962 & 17.9 & 42~~~ & 18.8 & TF & Ir & \\
PGC\,086673 & 112259.4+172826 & 1268 & 1719 & 17.72 & 60~~~ & 29.8 & TF & Ir & NGC\,3626 \\
AGC\,215414 & 112311.1+134220 & 746 & 1225 & 18.0 & 27~~~ & 11.1 & TF & Ir & NGC\,3627 \\
PGC\,1509123 & 112313.9+162711 & 1004 & 1463 & 18.37 & ~~~ & & & dE & NGC\,3607 \\
PGC\,034965 & 112318.8+035719 & 1415 & 1955 & 16.20 & ~~~ & & & dE & PGC\,34965 \\
IC\,2787 & 112319.1+133747 & 576 & 1055 & 15.70 & ~~~ & 10.3 & mem & dE & NGC\,3627 \\
IC\,2791 & 112337.6+125345 & 530 & 1014 & 17.15 & 22~~~ & 10.3 & mem & Ir & NGC\,3627 \\
KK~103 & 112341.1+191626 & 1789 & 2226 & 17.60 & ~~~ & & & Sph & \\
NGC\,3659 & 112345.4+174906 & 1173 & 1621 & 12.92 & ~~~ & 26.5 & tf & Sd & NGC\,3686 \\
SDSS\,J11240 & 112408.5+034404 & 1587 & 2128 & 18.13 & ~~~ & & & dE & \\
NGC\,3664 & 112424.3+031930 & 1203 & 1746 & 13.31 & ~~~ & & & Sm & NGC\,3640 \\
NGC\,3664A & 112425.0+031317 & 1147 & 1690 & 16.19 & 84~~~ & 26.3 & TF & Sm & NGC\,3640 \\
NGC\,3666 & 112426.1+112031 & 917 & 1411 & 12.70 & 255~~~ & 19.3 & tf & Sbc & \\
AGC\,215142 & 112444.5+151632 & 1008 & 1475 & 16.9 & 123~~~ & 39.1 & TFb & Sdm & NGC\,3607 \\
PGC\,035087 & 112501.8+170509 & 1093 & 1546 & 17.35 & ~~~ & & & Im & NGC\,3686 \\
AGC\,214317 & 112505.4+040716 & 1421 & 1959 & 17.83 & 130~~~ & & & Im & PGC\,34965 \\
PGC\,035096 & 112510.8+165304 & 903 & 1358 & 16.12 & ~~~ & & & Im & NGC\,3607 \\
PGC\,1502483 & 112529.0+161019 & 706 & 1166 & 16.88 & ~~~ & & & BCD & \\
AGC\,214318 & 112540.0+044036 & 1398 & 1933 & 18.06 & 123~~~ & & & BCD & \\
IC\,0692 & 112553.5+095914 & 1008 & 1510 & 14.23 & 95~~~ & 17.2 & TF & BCD & \\
AGC\,219119 & 112603.4+080432 & 1410 & 1924 & 18.7 & 35~~~ & 20.6 & TF & Ir & \\
AGC\,214319 & 112608.3+040345 & 1351 & 1889 & 17.24 & 49~~~ & 17.4 & TF & Im & \\
NGC\,3681 & 112629.8+165148 & 1124 & 1578 & 12.42 & ~~~ & & & Sab & NGC\,3686 \\
AGC\,215296 & 112655.2+145003 & 788 & 1257 & 19.12 & 44~~~ & & & Ir & NGC\,3627 \\
AGC\,219202 & 112710.9+050856 & 1348 & 1879 & 19.1 & 70~~~ & & & Ir & \\
IC\,2828 & 112711.0+084352 & 886 & 1396 & 15.03 & 80~~~ & 14.2 & TF & BCD & NGC\,3705 \\
NGC\,3684 & 112711.2+170149 & 1048 & 1501 & 12.31 & ~~~ & 23.8 & tf & Sbc & NGC\,3686 \\
AGC\,219203 & 112728.9+053702 & 1345 & 1873 & 18.8 & 28~~~ & & & Ir & \\
NGC\,3686 & 112743.9+171327 & 1043 & 1494 & 12.00 & ~~~ & 18.6 & tf & Sbc & NGC\,3686 \\
SDSS\,J11280 & 112806.2+175913 & 902 & 1347 & 18.19 & ~~~ & & & Im & NGC\,3686 \\
NGC\,3691 & 112809.4+165514 & 970 & 1423 & 12.64 & ~~~ & 28.3 & tf & Sbc & NGC\,3686 \\
NGC\,3692 & 112824.3+092427 & 1576 & 2081 & 13.14 & 408~~~ & 36.0 & tf & Sb & \\
AGC\,213939 & 112824.3+060704 & 1389 & 1914 & 17.55 & 45~~~ & 17.8 & TF & Ir & \\
KK~104 & 112851.2+181658 & 1200 & 1642 & 17.10 & 62~~~ & 23.7 & TF & Ir & NGC\,3686 \\
PGC\,1164263 & 112922.6+002220 & 1490 & 2046 & 18.69 & ~~~ & & & Ir & \\
PGC\,3123526 & 112930.0+031343 & 1329 & 1870 & 17.92 & ~~~ & & & BCD & \\
AGC\,213091 & 112934.6+104836 & 600 & 1095 & 17.74 & ~~~ & & & Ir & \\
PGC\,3287557 & 112945.6+003425 & 922 & 1476 & 18.94 & ~~~ & & & Ir & \\
PGC\,035426 & 112954.5+162546 & 951 & 1407 & 17.61 & ~~~ & & & Im & NGC\,3686 \\
NGC\,3705 & 113007.5+091636 & 868 & 1373 & 11.76 & 345~~~ & 18.4 & tf & Sab & NGC\,3705 \\
PGC\,3090344 & 113026.2+171957 & 1064 & 1513 & 16.41 & ~~~ & & & dEem? & NGC\,3686 \\
KKH~68 & 113053.3+140846 & 751 & 1223 & 17.8 & 22~~~ & 8.5 & TF & Ir & NGC\,3627 \\
AGC\,215303 & 113108.8+133414 & 875 & 1351 & 17.90 & 32~~~ & 13.0 & TF & Im & NGC\,3627 \\
Leo V & 113109.6+021312 & -7 & 538 & 17.6 & ~~~ & 0.18 & rgb & Sph & Milky Way \\
PGC\,035565 & 113201.9+143639 & 1002 & 1470 & 16.75 & 115~~~ & 39.6 & TF & Sdm & \\
PGC\,1228108 & 113244.1+022825 & 882 & 1425 & 17.45 & ~~~ & & & Ir & \\
PGC\,3291243 & 113306.9+012051 & 1466 & 2015 & 18.98 & ~~~ & & & Ir & \\
PGC\,1598409 & 113319.1+193551 & 1320 & 1750 & 17.23 & ~~~ & & & Im & \\
AGC\,215306 & 113350.1+144929 & 1007 & 1472 & 17.46 & 64~~~ & 29.0 & TF & Im & \\
AGC\,215248 & 113350.9+140315 & 808 & 1279 & 17.91 & 19~~~ & & & Im & NGC\,3627 \\
KK~107 & 113416.2+170947 & 941 & 1389 & 17.21 & ~~~ & & & Ir & \\
IC\,2934 & 113419.6+131917 & 1069 & 1545 & 15.6 & 61~~~ & 11.1 & TF & Im & \\
KKH~69 & 113453.3+110112 & 742 & 1233 & 17.6 & 22~~~ & 7.4 & TF & Ir & NGC\,3627 \\
SDSS\,J11345 & 113456.5+161452 & 1020 & 1474 & 17.87 & ~~~ & & & Ir & \\
AGC\,213169 & 113518.4+045717 & 1226 & 1754 & 18.4 & 37~~~ & 20.2 & TF & Ir & \\
PGC\,3291071 & 113530.2+015944 & 1440 & 1984 & 19.10 & ~~~ & & & Ir & \\
PGC\,1209232 & 113543.0+015325 & 1410 & 1954 & 18.31 & ~~~ & & & BCD & \\
UGC\,06578 & 113636.8+004858 & 915 & 1464 & 15.2 & 90~~~ & 18.1 & TF & Im & \\
AGC\,213155 & 113708.8+131504 & 855 & 1329 & 17.50 & 40~~~ & 14.8 & TF & Sd & NGC\,3810 \\
HIPASS\,J113 & 113728.7+182436 & 844 & 1280 & 18.5 & ~~~ & & & Ir & \\
UGC\,06594 & 113737.1+163322 & 925 & 1375 & 14.92 & ~~~ & 18.9 & tf & Scd & \\
NGC\,3773 & 113813.0+120644 & 849 & 1330 & 13.51 & 108~~~ & 17.1 & TF & BCD & NGC\,3810 \\
PGC\,1170677 & 113817.4+003648 & 773 & 1322 & 16.87 & ~~~ & & & Ir & \\
PGC\,1191771 & 113901.1+012012 & 1429 & 1974 & 16.89 & ~~~ & & & Im & \\
PGC\,1597887 & 113908.9+193500 & 1624 & 2050 & 18.59 & ~~~ & & & Ir & \\
IC\,0718 & 113952.8+085229 & 1836 & 2338 & 14.67 & 156~~~ & 27.9 & TF & Sdm & \\
IC\,0719 & 114018.5+090035 & 1686 & 2186 & 13.89 & 294~~~ & 28.6 & tf & S0 & \\
AGC\,215137 & 114056.7+140428 & 776 & 1242 & 16.5 & 110~~~ & 35.3 & TF & Scd & \\
NGC\,3810 & 114058.8+112816 & 857 & 1341 & 11.27 & 249~~~ & 18.3 & tf & Sc & NGC\,3810 \\
UGC\,06655 & 114150.6+155825 & 635 & 1087 & 14.93 & 54~~~ & & & BCD & \\
UGC\,06669 & 114218.7+145944 & 906 & 1365 & 16.65 & 65~~~ & 20.1 & TF & Im & \\
PGC\,1218595 & 114228.4+021050 & 1623 & 2161 & 18.57 & ~~~ & & & Sm & \\
UGC\,06670 & 114229.4+182000 & 820 & 1254 & 13.39 & 194~~~ & 18.1 & TF & Sd & \\
KDG~79 & 114310.5+141327 & 894 & 1358 & 17.4 & 85~~~ & 18.7 & TFb & Sm & NGC\,3810 \\
AGC\,213333 & 114327.0+112354 & 763 & 1246 & 16.71 & 64~~~ & 20.3 & TF & BCD & NGC\,3810 \\
PGC\,1519757 & 114440.7+165359 & 794 & 1237 & 17.23 & ~~~ & & & BCD & \\
KKH~72 & 114454.1+020951 & 839 & 1376 & 18.0 & 32~~~ & & & Ir & \\
SDSS\,J11470 & 114707.0+030623 & 844 & 1374 & 18.03 & ~~~ & & & Ir & \\
PGC\,2806928 & 114816.4+183833 & 879 & 1307 & 17.8 & ~~~ & & & Im & \\
SDSS\,J11484 & 114843.1+171053 & 988 & 1426 & 17.93 & ~~~ & & & Ir & \\
PGC\,1528400 & 114905.6+171521 & 519 & 957 & 18.49 & ~~~ & & & dE & \\
SDSS\,J11493 & 114931.0+151539 & 744 & 1196 & 17.8 & ~~~ & & & dE & \\
SDSS\,J11495 & 114957.1+161744 & 1080 & 1524 & 18.13 & ~~~ & & & Ir & \\
Mrk~750 & 115002.7+150124 & 635 & 1088 & 15.76 & 47~~~ & 13.6 & TF & BCD & \\
AGC\,210835 & 115055.9+143542 & 893 & 1349 & 16.66 & 52~~~ & 11.6 & TF & Im & \\
AGC\,213174 & 115104.8+051446 & 1324 & 1839 & 17.92 & ~~~ & & & BCD & \\
KIG\,0506 & 115201.9+135243 & 879 & 1339 & 15.79 & 79~~~ & 18.3 & TF & Sm & \\
SDSS\,J11530 & 115300.3+160230 & 826 & 1270 & 19.0 & ~~~ & & & Ir & \\
PGC\,166116 & 115401.6+164324 & 875 & 1313 & 17.37 & ~~~ & & & Ir & \\
IC\,0745 & 115412.3+000812 & 951 & 1491 & 14.11 & ~~~ & 18.3 & sbf & Eem & \\
AGC\,215145 & 115412.5+122606 & 880 & 1348 & 18.5 & 32~~~ & 17.1 & TF & Ir & \\
SDSS\,J11544 & 115449.3+064234 & 1209 & 1713 & 18.52 & ~~~ & & & Ir & \\
PGC\,3291881 & 115501.8+013900 & 1627 & 2159 & 18.82 & ~~~ & & & Ir & \\
KIG\,0511 & 115504.9+014311 & 1116 & 1647 & 15.5 & 112~~~ & 26.8 & TF & Sm & \\
UGC\,06903 & 115536.9+011414 & 1719 & 2252 & 14.09 & 177~~~ & & & Sc & \\
PGC\,135785 & 115722.4+014653 & 1811 & 2340 & 17.70 & ~~~ & & & Ir & \\
PGC\,1218832 & 115725.1+021116 & 839 & 1366 & 17.97 & ~~~ & & & BCD & PGC\,1218144 \\
PGC\,1218144 & 115735.2+021004 & 796 & 1323 & 16.49 & ~~~ & & & Sdm & PGC\,1218144 \\
PGC\,1488625 & 115840.4+153534 & 454 & 896 & 18.39 & ~~~ & & & Im & \\
AGC\,213178 & 115900.8+044011 & 1446 & 1958 & 17.20 & 58~~~ & 19.4 & TF & BCD & \\
PGC\,3291730 & 115909.2+012938 & 1989 & 2518 & 19.09 & ~~~ & & & Im & \\
AGC\,210968 & 115933.8+135315 & 1330 & 1784 & 15.2 & 57~~~ & 11.3 & TF & Sm & \\
\end{longtable}
\end{center}

\end{document}